\documentclass[a4paper]{report}
\usepackage[utf8]{inputenc}
\usepackage[T1]{fontenc}
\usepackage{RJournal}
\usepackage{amsmath,amssymb,array}
\usepackage{booktabs}
\usepackage{algorithm}
\usepackage{algpseudocode}
\usepackage{graphics}

\begin{document}

\sectionhead{Contributed research article}
\volume{XX}
\volnumber{YY}
\year{20ZZ}
\month{AAAA}

\begin{article}
\title{glmmPen: High Dimensional Penalized Generalized Linear Mixed Models}
\author{by Hillary M. Heiling, Naim U. Rashid, Quefeng Li, and Joseph G. Ibrahim}

\maketitle

\abstract{ 
Generalized linear mixed models (GLMMs) are widely used in research for their ability to model correlated outcomes with non-Gaussian conditional distributions. The proper selection of fixed and random effects is a critical part of the modeling process since model misspecification may lead to significant bias. However, the joint selection of fixed and random effects has historically been limited to lower-dimensional GLMMs, largely due to the use of criterion-based model selection strategies. Here we present the R package glmmPen, one of the first to select fixed and random effects in higher dimension using a penalized GLMM modeling framework. Model parameters are estimated using a Monte Carlo Expectation Conditional Minimization (MCECM) algorithm, which leverages Stan and RcppArmadillo for increased computational efficiency. Our package supports the Binomial, Gaussian, and Poisson families and multiple penalty functions. In this manuscript we discuss the modeling procedure, estimation scheme, and software implementation through application to a pancreatic cancer subtyping study. Simulation results show our method has good performance in selecting both the fixed and random effects in high dimensional GLMMs.
}
 


\section{Introduction}
\label{sec:introduction}

Generalized linear mixed models (GLMMs) are utilized in many disciplines, including the social sciences \citep{schmidt2016random}, biomedical sciences \citep{fitzmaurice2012applied}, public health and epidemiology \citep{szyszkowicz2006use,kleinman2004generalized,dean2007generalized}, natural sciences including ecology and evolution \citep{bolker2009generalized}, and economics \citep{langford1994using}. 
GLMMs are an extension of generalized linear models (GLMs) where the predictors within the model can have ``fixed'' or ``random'' effects. Coefficients corresponding to fixed effects predictors can be considered to describe population-level relationships between the predictors and the outcome. Random effects predictors pertain to variables whose relationships with the outcome are presumed to vary randomly across ``groups'' of observations within the data, leading to group-specific coefficient estimates \citep{fitzmaurice2012applied}.
In practical applications, these ``groups'' may pertain to clusters of samples, repeated measures within the same individual, or observations resulting from nested designs.  Multiple studies have shown that omitting important random effects can lead to bias in the estimated variance of the fixed effects; conversely, including unnecessary random effects may lead to computational difficulties \citep{thompson2017bias, gurka2011avoiding, bondell2010joint}. As a result, proper specification of fixed and random effects is a critical step in the application of GLMMs.  

In many low dimensional settings, researchers may have \textit{a priori} knowledge about which variables are fixed or random. For instance, researchers may reasonably expect treatment effects in multi-site clinical trials to vary by site \citep{feaster2011modeling}. However, in high dimensional settings, it is often not known \textit{a priori} which variables should be specified as fixed or random in the model. In such settings, the feature space may also be sparse, with many variables unrelated to the outcome.  Therefore, variable selection approaches are employed to evaluate and select from a set of candidate models.   R packages such as \CRANpkg{lme4} \citep{lme42007}, \CRANpkg{mcemGLM} \citep{mcemglm2015}, and \CRANpkg{MCMCglmm} \citep{mcmcglmm2010} allow users to fit a pre-specified set of models, which may then be compared using model selection criteria such as the profile conditional AIC \citep{donohue2011conditional}, the BIC-ICQ criterion \citep{BICq2011}, the hybrid Bayesian information criterion, BICh \citep{BICh2014}, or other criteria developed for mixed effects models. However, criterion-based all-subsets selection or direct model comparison strategies are not feasible even in small dimensions, as with $p$ predictors there are $2^{2p}$ possible combinations of fixed and random effects to be evaluated. 

Packages such as \CRANpkg{glmnet} \citep{glmnet2010}, \CRANpkg{ncvreg} \citep{ncvreg2011}, and \CRANpkg{grpreg} \citep{grpreg2015} avoid this limitation for GLMs via coordinate-descent based penalized likelihood methods for variable selection, and therefore scale much better with respect to $p$.  Unfortunately, none of these methods can account for random effects in their variable selection procedure. Other packages such as \CRANpkg{glmmLasso} \citep{glmmLassoManual} and \pkg{glmmixedLASSO} \citep{schelldorfer2014glmmLASSO} alternatively allow the inclusion of random effects in the model while performing variable selection, but only allow for variable selection on the fixed effects. Prior work has shown that simultaneous selection of fixed and random effects is desirable because improper specification of the random effects can significantly affect the selection of the fixed effects, and vice versa \citep{bondell2010joint}. In addition, there may not be \textit{a priori} knowledge of which variables have effects that vary randomly across groups.
Therefore, the specification of random effects may be difficult in practical applications, particularly as the dimension of the data grows.

To address these limitations in performing variable selection in high-dimensional GLMMs, we present the \CRANpkg{glmmPen} R package. This package allows for the simultaneous selection of fixed and random effects predictors in higher dimensions through the use of penalized generalized linear mixed models (pGLMMs). Similar to \CRANpkg{ncvreg} and \CRANpkg{glmnet}, this package focuses on variable selection for the purpose of creating prediction models, and does not provide methods for statistical inference.
The package leverages Monte Carlo Expectation Conditional Minimization (MCECM) in combination with several techniques to improve the computational efficiency of the  algorithm. In the MCECM E-step, \CRANpkg{glmmPen} utilizes the \pkg{Stan} software implemented in the \CRANpkg{rstan} package to efficiently sample from the posterior distribution of the random effects, and a Majorization-Minimization coordinate descent algorithm is utilized to update model parameters in the M-step. The \CRANpkg{glmmPen} package utilizes the fast looping capabilities within \CRANpkg{Rcpp} and \CRANpkg{RcppArmadillo} in order to recalculate large matrices within intermediate computing steps without needing to store them, improving memory use. The \CRANpkg{glmmPen} package is also able to improve the speed of the overall variable selection procedure by strategic coefficient initialization (see Section ``Initialization and convergence'') and strategic restriction of random effects (see Section ``Tuning parameter selection strategy''). 

The main estimation functions of the package are  \code{glmmPen} and \code{glmm}, where the latter can be used to fit traditional generalized linear mixed models without penalization. The user interface and output of the \code{glmmPen} and \code{glmm} functions were designed to be very similar to those from the functions \code{lmer} and \code{glmer} to facilitate ease of use. Specifically, \CRANpkg{glmmPen} outputs a \code{pglmmObj} object which, like the \code{merMod} object from \CRANpkg{lme4}, can facilitate the application of common S3 method functions used by \CRANpkg{lme4} such as \code{logLik}, \code{fixef}, \code{ranef}, and others. In addition, multiple types of penalties and information criteria for selecting optimal penalties are available in the package, and the package supports the Binomial, Gaussian, and Poisson distributional families.

Our manuscript is organized as follows. We begin in Section 2 by reviewing the pGLMMs modeling framework, first described in \cite{rashid2020}. Section 3 describes the MCECM algorithm used by \CRANpkg{glmmPen} to fit pGLMM models. Section 4 describes the variable selection procedure of the package and the Bayesian information criterion (BIC) type selection criteria available for use. Section 5 illustrates a practical application of the \CRANpkg{glmmPen} R package using data from a recent cancer subtyping study. Section 6 provides some simulation results. Finally, we provide concluding comments in Section 7. The package is available from the Comprehensive R Archive Network (CRAN) at \url{https://cran.r-project.org/package=glmmPen}. The replication of all code content, tables, and figures presented in this paper can be found in the GitHub repository \url{https://github.com/hheiling/paper_glmmPen_RJournal}. Supplementary results mentioned but not reported in this paper can also be found in this GitHub repository.

\section{Generalized linear mixed models}
\label{sec:modelinfo}

We review the notation and model formulation of our approach, first introduced in  \cite{rashid2020}. We consider the case where we want to analyze data from \(K\) independent groups of any kind. For instance, we could be interested in analyzing data from \(K\) different studies, or longitudinal data from \(K\) individuals. For each group \(k = 1,...,K\), there are \(n_k\) observations for a total sample size
of \(N = \sum_{k=1}^K n_k\). For the \(k^{th}\) group, let \(\boldsymbol y_k = (y_{k1},...,y_{kn_k})^\top\) be the vector of \(n_k\) independent responses, let
\(\boldsymbol x_{ki} = (x_{ki,1},...,x_{ki,p})^\top\) be the \(p\)-dimensional vector of predictors, and let
\(\boldsymbol X_k = (\boldsymbol x_{k1}, ..., \boldsymbol x_{kn_k})^\top\). Although the \CRANpkg{glmmPen} package allows for different \(n_k\) for the
\(K\) groups, we will set \(\{n_k\}_{k=1}^K = n\) to simplify the notation within the equations presented in this paper. In GLMMs, we
assume that the conditional distribution of \(\boldsymbol y_k\) given \(\boldsymbol X_k\) belongs to the exponential family and has the
following density: \begin{equation}
  f(\boldsymbol y_k | \boldsymbol X_k, \boldsymbol \alpha_k; \theta) = 
    \prod_{i=1}^{n} c(y_{ki}) \exp[\tau^{-1} \{y_{ki} \eta_{ki} - b(\eta_{ki})\}],
    \label{eqn:exp_family}
\end{equation} where \(c(y_{ki})\) is a constant that only depends on
\(y_{ki}\), \(\tau\) is the dispersion parameter, \(b(\cdot)\) is a known
link function, and \(\eta_{ki}\) is the linear predictor. The \CRANpkg{glmmPen} algorithm currently allows for the Gaussian, Binomial, and
Poisson families with canonical links.

In the GLMM, the linear predictor has the form
\begin{equation}
  \eta_{ki} = \boldsymbol x_{ki}^\top \boldsymbol \beta + \boldsymbol z_{ki}^\top \boldsymbol \Gamma \boldsymbol \alpha_k,
  \label{eqn:linpred}
\end{equation} where \(\boldsymbol \beta = (\beta_1,...,\beta_{p})^\top\)
is a \(p\)-dimensional vector for the fixed effects coefficients (including the intercept),
\(\boldsymbol \alpha_k\) is a \(q\)-dimensional vector of unobservable
random effects (including the random intercept), \(\boldsymbol z_{ki}\) is a \(q\)-dimensional
subvector of \(\boldsymbol x_{ki}\), and \(\boldsymbol \Gamma\) is a
lower triangular matrix. In this notation, $\boldsymbol z_{ki}$ represents the random effects predictors, i.e. the subset of the total predictors ($\boldsymbol x_{ki}$) that have predictor effects that randomly vary across levels of the grouping variable.

In \cite{rashid2020}, the random effects vector \(\boldsymbol \alpha_k\)
is assumed to follow \(N_q(\boldsymbol 0, \boldsymbol I)\)
so that
\(\boldsymbol \Gamma \boldsymbol \alpha_k\) follows \(N(\boldsymbol 0, \boldsymbol{\Gamma \Gamma}^\top)\).
In this way, the random component of the linear predictor has variance
Var(\(\boldsymbol \Gamma \boldsymbol \alpha_k\)) =
\(\boldsymbol{\Gamma \Gamma}^\top\).

To simplify the procedure of estimating \(\boldsymbol \Gamma\), we
consider a vector \(\boldsymbol \gamma\) containing all of the nonzero
elements of \(\boldsymbol \Gamma\) such that \(\boldsymbol \gamma_t\) is
a \(t\) x 1 vector consisting of nonzero elements of the \(t^{th}\) row
of \(\boldsymbol \Gamma\) and
\(\boldsymbol \gamma = (\boldsymbol \gamma_1^\top,...,\boldsymbol \gamma_{q}^\top)^\top\).
We can then reparameterize the linear predictor
\citep{chen2003, BICq2011} to 
\begin{equation}
  \eta_{ki} = \boldsymbol x_{ki}^\top \boldsymbol\beta + \boldsymbol z_{ki}^\top \boldsymbol\Gamma \boldsymbol\alpha_k = \left (\boldsymbol x_{ki}^\top \hspace{10 pt} (\boldsymbol\alpha_k \otimes \boldsymbol z_{ki})^\top \boldsymbol J_q \right)
  \left (
  \begin{matrix}
    \boldsymbol\beta \\ \boldsymbol\gamma 
  \end{matrix}
  \right )
  \label{eqn:linpredJ}
\end{equation} where \(\boldsymbol J_q\) is a matrix that transforms
\(\boldsymbol \gamma\) to vec(\(\boldsymbol \Gamma\)) such that
vec(\(\boldsymbol \Gamma\)) = \(\boldsymbol J_q \boldsymbol \gamma\). \(\boldsymbol J_q\) is of
dimension \(q^2 \times q(q + 1)/2\) when the random effects
covariance matrix \(\boldsymbol{\Gamma \Gamma}^\top\) is unstructured;
alternatively, \(\boldsymbol J_q\) is of dimension \(q^2 \times q\) when the
random effects covariance matrix has an independence structure
(i.e.,~diagonal). The vector of parameters
\(\boldsymbol \theta = (\boldsymbol \beta^\top, \boldsymbol \gamma^\top, \tau)^\top\)
are the main parameters of interest. We denote the true value of
\(\boldsymbol \theta\) as
\(\boldsymbol \theta^{*} = (\boldsymbol \beta^{*\top}, \boldsymbol \gamma^{*\top}, \tau^{*})^\top = \text{argmin}_{\boldsymbol \theta}\text{E}_{\boldsymbol\theta}[-\ell(\boldsymbol \theta)]\)
where \(\ell(\boldsymbol \theta)\) is the observed marginal
log-likelihood across all \(K\) groups such that
\(\ell(\boldsymbol \theta) = \sum_{k=1}^K \ell_k(\boldsymbol \theta)\),
\(\ell_k(\boldsymbol \theta) = (1/n) \log \int f(\boldsymbol y_k | \boldsymbol X_k, \boldsymbol \alpha_k; \boldsymbol \theta) \phi(\boldsymbol \alpha_k) d \boldsymbol \alpha_k\).

Let us consider the high dimensional case where we want to select the
true nonzero fixed effects and true nonzero random effects. In other words, we
aim to identify the set \begin{equation*}
  S = S_1 \cup S_2 = \{j: \beta_j^* \ne 0 \} \cup \{t: ||\boldsymbol\gamma_t^*||_2 \ne 0\},
\end{equation*} where the set \(S_1\) represents the selection of true
nonzero fixed effects and the set \(S_2\) represents the selection of true
nonzero random effects. When \(\boldsymbol \gamma_t = \boldsymbol 0\),
this sets row \(t\) of \(\boldsymbol \Gamma\) entirely equal to 0,
indicating that effect of covariate \(t\) is fixed across the \(K\)
groups.

We aim to solve the following penalized likelihood: \begin{equation}
  \widehat{\boldsymbol\theta} = \text{argmin}_{\boldsymbol\theta} - \ell (\boldsymbol\theta) + \lambda_0 \sum_{j=1}^{p} \rho_0 \left (\beta_j \right ) + \lambda_1 \sum_{t=1}^{q} \rho_1 \left (||\boldsymbol\gamma_t||_2 \right ),
  \label{eqn:penlik}
\end{equation} where \(\ell(\boldsymbol \theta)\) is the observed
marginal log-likelihood for all \(K\) groups defined earlier,
\(\rho_0(t)\) and \(\rho_1(t)\) are general folded-concave penalty
functions, and \(\lambda_0\) and \(\lambda_1\) are positive tuning
parameters. In the \CRANpkg{glmmPen} package, the \(\rho_0(t)\) penalty
function options include the least absolute shrinkage and selection operator (LASSO) \(L_1\) penalty, the minimax concave penalty (MCP), and the smoothly clipped absolute deviation (SCAD) penalty \citep{glmnet2010, ncvreg2011}. For the
\(\rho_1(t)\) penalty, we treat the elements of \(\boldsymbol \gamma_t\)
as a group and penalize them in a groupwise manner using the group
LASSO, group MCP, or group SCAD penalties presented by Breheny and Huang
\citeyearpar{grpreg2015}. These groups of \(\boldsymbol \gamma_t\) are
then estimated to be either all zero or all nonzero. In this way, we
select covariates to have varying effects
(\(\boldsymbol{\widehat \gamma_t} \ne \boldsymbol 0\)) or fixed effects
(\(\boldsymbol{\widehat \gamma_t} = \boldsymbol 0\)) across the \(K\)
groups.

Similar to other variable selection packages such as package \CRANpkg{ncvreg}
\citep{ncvreg2011}, in \CRANpkg{glmmPen} we standardize the fixed effects covariates matrix
\(\boldsymbol X = (\boldsymbol X_1^\top,...,\boldsymbol X_K^\top)^\top\) such
that \\ \(\sum_{k=1}^K \sum_{i=1}^{n_{k}} x_{ki,j} = 0\) and
\(N^{-1} \sum_{k=1}^K \sum_{i=1}^{n_{k}} x_{ki,j}^2 = 1\) for
\(j = 1,...,p\); this process is performed automatically within the algorithm. Although the package \CRANpkg{grpreg} \citep{grpreg2015}
orthogonalizes grouped effects, we have found through simulations during early package testing that
first standardizing the fixed effects and then using subsets of these
standardized fixed effects for the random effects (recall:
\(\boldsymbol z_{ki}\) is a \(q\)-dimensional subvector of
\(\boldsymbol x_{ki}\)) is sufficient. During the selection procedure,
the fixed effects intercept and the variance of the random effects intercept remain unpenalized.

\section{MCECM algorithm}
\label{sec:mcecm}

We solve Equation~\ref{eqn:penlik} for a specific $(\lambda_0,\lambda_1)$ penalty parameter combination using a Monte Carlo Expectation
Conditional Minimization (MCECM) algorithm \citep{garcia2010}. The MCECM algorithm described in this section uses many of the steps and assumptions described in \citet{rashid2020}, but here we provide further practical details about the E-step, M-step, initialization, and convergence. Additionally, the implementation outlined in this paper has several improvements to the implementation used in \citet{rashid2020}. In \CRANpkg{glmmPen}, the E-step allows for several possible sampling schemes, including the fast and efficient No-U-Turn Hamiltonian Monte Carlo sampling procedure (NUTS HMC) from the \pkg{Stan} software \citep{stan2017, hoffman2014NUTS}. The \CRANpkg{glmmPen} package was also able to reduce the required memory usage of the MCECM algorithm. In the M-step, we utilized the fast looping capability of packages \CRANpkg{Rcpp} and \CRANpkg{RcppArmadillo} to allow for fast recalculation of large matrices (see Step 3 of the M-step presented in Algorithm \ref{alg:mstep}) and avoid their storage, improving model scalability.

During the MCECM algorithm, we  aim  to  evaluate  (E-step)  and  minimize  (M-step)  the following penalized Q-function in the \(s^{th}\) iteration of the algorithm: \begin{align}
  \begin{aligned}
    Q_\lambda(\boldsymbol\theta | \boldsymbol \theta^{(s)}) & = \sum_{k=1}^K E \left \{ -\log(f(\boldsymbol y_k, \boldsymbol X_k, \boldsymbol\alpha_k;\boldsymbol\theta | \boldsymbol D_o; \boldsymbol\theta^{(s)})) \right \} + \lambda_0 \sum_{j=1}^{p} \rho_0 \left (\beta_j \right ) + \lambda_1 \sum_{t=1}^{q} \rho_1 \left (||\boldsymbol\gamma_t||_2 \right ) \\
    & = Q_1(\boldsymbol\theta | \boldsymbol\theta^{(s)}) + Q_2(\boldsymbol\theta^{(s)}) + \lambda_0 \sum_{j=1}^{p} \rho_0 \left (\beta_j \right ) + \lambda_1 \sum_{t=1}^{q} \rho_1 \left (||\boldsymbol\gamma_t||_2 \right ),
    \label{eqn:Qfun}
  \end{aligned}
\end{align} where
\((\boldsymbol y_k, \boldsymbol X_k, \boldsymbol \alpha_k)\) gives the
complete data for group $k$,  \(\boldsymbol D_{k,o} = (\boldsymbol y_k, \boldsymbol X_k)\) gives the observed data
for group $k$, and \(\boldsymbol D_o\) represents the entirety of the observed
data. In other words, we aim to evaluate and minimize the penalized expectation of the negative joint log-likelihood with respect to the observed data. From \cite{rashid2020}, the expectation can be written as the sum of the following terms:
\begin{equation}
  Q_1(\boldsymbol\theta | \boldsymbol\theta^{(s)}) = - \sum_{k=1}^K \int \log [f(\boldsymbol y_k | \boldsymbol X_k, \boldsymbol\alpha_k; \boldsymbol\theta)] \phi(\boldsymbol\alpha_k | \boldsymbol D_{k,o}; \boldsymbol\theta^{(s)}) d \boldsymbol\alpha_k,
  \label{eqn:Q1}
\end{equation} \begin{equation}
   Q_2(\boldsymbol\theta^{(s)}) = - \sum_{k=1}^K \int \log [\phi(\boldsymbol{\alpha}_k)] \phi(\boldsymbol\alpha_k | \boldsymbol D_{k,o}; \boldsymbol\theta^{(s)}) d \boldsymbol\alpha_k
  \label{eqn:Q2}
\end{equation}
The $Q_1(\boldsymbol\theta | \boldsymbol\theta^{(s)})$ function expresses the conditional model of the observed data given the latent (random) variables and integrates over the latent variables. Using the $Q_1(\boldsymbol\theta | \boldsymbol\theta^{(s)})$ function, we aim to derive the fixed and random effect coefficient estimates during the M-step of the algorithm. During the E-step, we aim to approximate the integral in the $Q_1(\boldsymbol\theta | \boldsymbol\theta^{(s)})$ function by incorporating samples from the posterior distribution of the latent variables.

\subsection{Monte Carlo E-step}
\label{sec:estep}
The integrals in the Q-function do not have closed forms when $f(\boldsymbol y_k | \boldsymbol X_k, \boldsymbol\alpha_k^{(s,m)}; \boldsymbol\theta)$ is assumed to be non-Gaussian, and become difficult to approximate as $q$ (the number of random effect predictors) increases. Consequently, we approximate these integrals using a Markov chain Monte Carlo (MCMC) sample of size M from the posterior density
\(\phi(\boldsymbol \alpha_k | \boldsymbol D_{k,o}; \boldsymbol \theta^{(s)})\).
The \CRANpkg{glmmPen} package can draw samples from this posterior using one of
several techniques: the No-U-Turn Hamiltonian Monte Carlo sampling
procedure (NUTS HMC) implemented by the \pkg{Stan} software, which \CRANpkg{glmmPen} calls
using the \CRANpkg{rstan} package (\cite{stan2017}; default, and strongly
recommended for its speed and efficiency); Metropolis-within-Gibbs with
an adaptive random walk sampler \citep{adaptMCMC2009}; and
Metropolis-within-Gibbs with an independence sampler
\citep{compstats2012}. Each sampler type uses a standard normal
candidate distribution. Let \(\boldsymbol \alpha_k^{(s,m)}\) be the
\(m^{th}\) simulated value, \(m = 1,...,M\), at the \(s^{th}\) iteration
of the algorithm for group \(k\). The integral in Equation~\ref{eqn:Q1} can then be approximated as \begin{equation*}
  Q_1(\boldsymbol\theta | \boldsymbol\theta^{(s)}) \approx - \frac{1}{M} \sum_{m=1}^M \sum_{k=1}^K \log f(\boldsymbol y_k | \boldsymbol X_k, \boldsymbol\alpha_k^{(s,m)}; \boldsymbol\theta).
\end{equation*}

Although the optimal number of MCMC samples  \(M^{(s)}\) in the E-step at EM iteration $s$ 
is not well defined, the general consensus is that a smaller sample size of the posterior is suitable for the start of the algorithm but larger sample sizes are needed later in the algorithm \citep{booth1999}. We set the default number of MCMC samples in the first iteration of the MCECM algorithm \(M^{(1)} = 250\) when \(q \leq 10\), and \(M^{(1)}= 100\) otherwise (we decrease the initial sampling size when the number of random effects predictors is large in order to help speed up the algorithm). Then, in a manner similar to the \CRANpkg{mcemGLM} package \citep{mcemglm2015},
the MCMC sample size is increased by a multiplicative factor \(v\) at
each step of the algorithm such that \(M^{(s)} = v \times M^{(s-1)}\) until
either the value of \(M^{(s)}\) reaches its maximum allowed value or the
EM algorithm converges. In \CRANpkg{glmmPen}, the default maximum allowed value is dependent on the the number of random effects in the model, $q$ (see the documentation of \code{optimControl} for more details). For the first 15
iterations of the EM algorithm, the value of \(v\) is set to 1.1. For the
remaining steps of the algorithm, \(v\) is set to 1.2.

\subsection{M-step}
\label{sec:mstep}
In the M-step of the algorithm, we aim to minimize \begin{equation}
  Q_{1,\lambda}(\boldsymbol\theta | \boldsymbol\theta^{(s)}) = Q_1(\boldsymbol\theta | \boldsymbol\theta^{(s)}) + \lambda_0 \sum_{j=1}^{p} \rho_0 \left (\beta_j \right ) + \lambda_1 \sum_{t=1}^{q} \rho_1 \left (||\boldsymbol\gamma_t||_2 \right )
  \label{eqn:Mstep}
\end{equation} with respect to
\(\boldsymbol \theta = (\boldsymbol \beta^\top, \boldsymbol \gamma^\top, \tau)^\top\).
The minimization of Equation~\ref{eqn:Mstep} with respect to
\(\boldsymbol \beta\) and \(\boldsymbol \gamma\) is performed using a
Majorization-Minimization approach. For the general
exponential family, Rashid et al. \citeyearpar{rashid2020} suggested
minimizing with respect to \(\tau\) using the standard optimization
algorithm Newton-Raphson. In \CRANpkg{glmmPen}, the only family implemented
with a dispersion parameter is the Gaussian family, and the variance
\(\sigma^2\) can be estimated directly from a derivation of the Q
function conditional on the most recent updates of
\(\boldsymbol \beta^{(s)}\) and \(\boldsymbol \gamma^{(s)}\):
\begin{equation}
  \sigma^2 = \frac{1}{M\times N} \sum_{m=1}^M \sum_{k=1}^K \sum_{i=1}^{n_k} (y_{ki} - \eta_{ki}^{(s,m)})^2,
  \label{eqn:sigma}
\end{equation} where \(\eta_{ki}^{(s,m)}\) is the linear predictor
\(\eta_{ki}\) evaluated with \(\boldsymbol \beta^{(s)}\),
\(\boldsymbol \gamma^{(s)}\), and sample
\(\boldsymbol \alpha_k^{(s,m)}\).

Let $s$ represent the iteration of the MCECM algorithm, and $h$ represent the iteration within a particular M-step of the MCECM algorithm. The M-step of the \(s^{th}\) iteration of the MCECM algorithm proceeds
as in Algorithm \ref{alg:mstep}.

\begin{algorithm}[h!]
\caption{M-step of the $s$-th iteration of the MCECM algorithm} \label{alg:mstep}
\begin{algorithmic}
\State 1. Coefficient parameter estimates from the previous M-step, $\boldsymbol \theta^{(s-1)}$, are used to initialize the coefficient parameters of the current M-step at M-step iteration $h=0$, denoted $\boldsymbol \theta^{(s,0)}$. \\
\State 2. Conditional on $\boldsymbol \gamma^{(s,h-1)}$ and $\tau^{(s-1)}$, each $\beta_j^{(s,h)}$ for $j = \{1,...,p\}$ is given a single update using the Majorization-Minimization algorithm specified by Breheny and Huang (2015). \\
\State 3. For each group k in $k = \{1,...,K\}$, the augmented matrix $\boldsymbol{\tilde z}_{ki} = (\boldsymbol{\tilde \alpha}_k^{(s)} \otimes \boldsymbol z_{ki}) J_q$ is created for $i = 1,...,n_k$ where $\boldsymbol{\tilde \alpha}_k^{(s)} = ((\boldsymbol \alpha_k^{(s,1)})^\top,...,(\boldsymbol \alpha_k^{(s,M)})^\top)^\top$. This augmented matrix is used in the random effect portion of the linear predictor specified in Equation~\ref{eqn:linpredJ}. The dimension of $\boldsymbol{\tilde z}_{ki}$ is $M \times q(q+1)/2$ for an unstructured covariance matrix and $M \times q$ for an independent covariance matrix. This augmented matrix is used to calculate Equation 2.9 in \cite{grpreg2015}. \\
\State 4. Conditional on the $\tau^{(s-1)}$ and the recently updated $\boldsymbol \beta^{(s,h)}$, each $\boldsymbol \gamma_t^{(s,h)}$ for $t = \{1,...,q\}$ is updated using the Majorization-Minimzation coordinate descent grouped variable selection algorithm specified by \cite{grpreg2015}, except the residuals are not updated after every $\boldsymbol \gamma_t^{(s,h)}$ coefficient update. \\
\State 5. Steps 2 through 4 are repeated until the M-step convergence criteria specified in Equation~\ref{eqn:conv_mstep} are reached or until the M-step reaches its maximum number of iterations: 
  \begin{equation}
    \max \left \{ \max_j |\beta_j^{(s,h+1)} - \beta_j^{(s,h)}|, \max_{t,l} |\gamma_{tl}^{(s,h+1)} - \gamma_{tl}^{(s,h)}| \right \} < \delta,
    \label{eqn:conv_mstep}
  \end{equation}
where $\gamma_{tl}$ is an individual element of $\boldsymbol \gamma_t$. The default value of $\delta$ is 0.0005.\\
\State 6. Conditioning on the newly updated $\boldsymbol \beta^{(s)}$ and $\boldsymbol b^{(s)}$, $\tau^{(s)}$ is updated (generically, using the Newton-Raphson algorithm; for Gaussian family, using Equation~\ref{eqn:sigma}).
\end{algorithmic}
\end{algorithm}

Algorithm \ref{alg:mstep} recomputes the augmented matrices
\(\boldsymbol{\tilde z}_{ki}\) for \(k=1,...,K\) and \(i=1,...,n_k\) in
step 3 of every M-step iteration \(h\) for several reasons. These repeat
calculations prevent the M-step from having to store the augmented matrix
\(\boldsymbol{\tilde Z} = (\boldsymbol{\tilde Z}_1^\top,...,\boldsymbol{\tilde Z}_K^\top)^\top\)
where
\(\boldsymbol{\tilde Z}_k = (\boldsymbol{\tilde z}_{k1}^\top,...,\boldsymbol{\tilde z}_{kn_k}^\top)^\top\).
This full augmented matrix is of dimension
\((M\times N) \times q(q + 1)/2\) or \((M\times N) \times q\) depending on
whether the random effect covariance matrix is unstructured or
independent, respectively. As the MCMC sample size increases throughout
the MCECM algorithm and as \(q\) increases, saving this
\(\boldsymbol{\tilde Z}\) becomes more and more memory prohibitive even
when utilizing large matrix implementation tools such as the package
\CRANpkg{bigmemory} \citep{bigmemory2013}. During testing, we
found that recomputing the \(\boldsymbol{\tilde z}_{ki}\) matrices
during each M-step iteration utilizing \CRANpkg{Rcpp} \citep{rcpp2011} and
\CRANpkg{RcppArmadillo} \citep{rcpparmadillo2014} significantly reduced the
time and memory required to compute each M-step.

In step 4 of the M-step, the residuals are not updated after every update to the random effects coefficients
\(\boldsymbol \gamma_t^{(s,h)}\) for \(t=1,...,q\) in order to speed up computation. Otherwise, 
this would require re-calculation of the augmented matrix specified in step 3 for each of the $q$ random effects within each M-step iteration. When \(q\) is large, this makes the M-step prohibitively time-consuming. Based on early package testing, simplifying step 4 with no residual updates speeds up the computation time in high dimensional settings and was found to have negligible impact on estimation accuracy.

The full MCECM algorithm then proceeds in Algorithm \ref{alg:MCECM}.

\begin{algorithm}[h!]
\caption{Full MCECM algorithm for single $(\lambda_0,\lambda_1)$ penalty combination} \label{alg:MCECM}
\begin{algorithmic}
  \State 1. Fixed and random effects $\boldsymbol \beta^{(0)}$ and $\boldsymbol \gamma^{(0)}$ are initialized as discussed in Section ``Initialization and convergence''. \\
  \State 2. E-step: In each E-step for EM iteration $s$, a burn-in sample from the posterior distribution of the random effects is run and discarded.
  A sample of size $M^{(s)}$ from the posterior is then drawn and retained for the M-step (see Section ``Monte Carlo E-step'' for details on default burn-in sample size, default $M^{(s)}$, and other E-step details). \\
  \State 3. M-step: Parameter estimates $\boldsymbol \beta^{(s)}$, $\boldsymbol \gamma^{(s)}$, and $\tau^{(s)}$ are then updated as described in Algorithm \ref{alg:mstep}. \\
  \State 4. Steps 2 and 3 are repeated until the average Euclidean distance between the vector containing the current coefficients $\boldsymbol \beta^{(s)}$ and $\boldsymbol \gamma^{(s)}$ and the vector containing the coefficients from $t$ EM iterations prior (default $t=2$) is less than $\epsilon$ (default $\epsilon=0.0015$) for at least two consecutive EM iterations or until the maximum number of EM iterations is reached (see Section ``Initialization and convergence'' for additional details). \\
  \item 5. Using the estimates of $\boldsymbol \beta$, $\boldsymbol \gamma$, and $\tau$ at EM convergence, a final sample from the posterior distribution of the random effects is drawn for use in the calculation of the marginal log-likelihood as well as for diagnostics of the MCMC chain. The marginal log-likelihood is used for model selection and is discussed in detail in Section \ref{sec:model-selection}.
\end{algorithmic}
\end{algorithm}

\subsection{Initialization and convergence}
\label{sec:initial}
The initial values of the fixed effects $\boldsymbol \beta^{(0)}$ and the Cholesky decomposition of the random effects covariance matrix $\boldsymbol \gamma^{(0)}$ for MCECM iteration $s=0$ are chosen in one of two ways. We discuss first the initialization procedure used when the package \CRANpkg{glmmPen} is used to fit a single model (\code{glmm} function) or the first model in the sequence of models fit for variable selection (\code{glmmPen} function). In this scenario, the fixed effects $\boldsymbol \beta^{(0)}$ are initialized by fitting a `naive' model using the coordinate descent techniques of Breheny and Huang (2011) assuming no random effects, and the random effects covariance matrix is initialized as a diagonal matrix with positive variance. This approach is similar to the \CRANpkg{mcemGLM} package. 

By default, this starting variance is initialized in an automated fashion. First, a GLMM composed of only a fixed and random intercept is fit using the \CRANpkg{lme4} package. The random intercept variance from this model is then multiplied by 2, and this value is set as the starting values of the diagonal of the random effects covariance matrix. We use this approach so that the starting variance of the random effects is sufficiently large, which helps improve the stability of the algorithm \citep{misztal2008reliable}. 


The MCMC chain used in the E-step of the algorithm, which approximately samples from the posterior density $\phi(\boldsymbol \alpha_k|\boldsymbol D_{k,o}; \boldsymbol \theta^{(s)})$ for groups $k=\{1,...,K\}$, is initialized in iteration $s=1$ with draws from the standard normal distribution. For all following iterations $s>1$, the MCMC chain is initialized with the last draw from the previous EM iteration $s-1$. 

When the algorithm performs variable selection using the \code{glmmPen} function, the model pertaining to the first tuning parameter combination evaluated is initialized using approach described above. For all subsequent tuning parameter combinations evaluated in the sequence, the fixed effects, random effects covariance matrix, and random effects MCMC chain are initialized using results from the previous tuning parameter fit. More details about initialization for variable selection is discussed in Section ``Tuning parameter selection''.

The EM algorithm runs until is the algorithm converges, defined as meeting the condition given in Equation~\ref{eqn:conv_EM} at least 2 consecutive times (default), or until the maximum number of EM iterations is reached:
\begin{equation}
    ||(\boldsymbol\beta^{(s)\top}, \boldsymbol\gamma^{(s)\top})^\top - (\boldsymbol\beta^{(s-t)\top}, \boldsymbol\gamma^{(s-t)\top})^\top||_2^2 / c^{s-t} < \epsilon
    \label{eqn:conv_EM}
  \end{equation}
where the superscript $(s-t)$ indicates the EM iteration $t$ iterations prior, $||.||_2^2$ represents the $L_2$ norm, and $c^{s-t}$ equals the total number of non-zero $(\boldsymbol \beta^\top, \boldsymbol \gamma^\top)^\top$ coefficients in iteration $(s-t)$. In other words, the algorithm computes the average Euclidean distance between the current coefficient vector $(\boldsymbol \beta^\top, \boldsymbol \gamma^\top)^\top$ and the coefficient vector from $t$ EM iterations prior (default $t=2$) and compares it with $\epsilon$, which has a default value of 0.0015.

This MCECM algorithm is able to handle much larger dimensions of $p$ fixed effect predictors and $q$ random effect predictors relative to prior methods for simultaneous fixed and random effects variable selection \citep{bondell2010joint,BICq2011}.  When the number of random effect predictors is greater than or equal to 10, we recommend approximating the random effect covariance
matrix \(\boldsymbol{\Gamma \Gamma}^\top\) as a diagonal matrix. In the mixed model setting, Fan and Li \citeyearpar{fanli2012} demonstrated both theoretical and empirical advantages to estimating the random effects covariance matrix in this manner as the number of random effect predictors $q$ grows. Empirically, they found that this approximation had a relatively low impact on the overall bias of the coefficients and resulted in a relatively large reduction of accumulated estimation error since many fewer covariance parameters needed to be estimated. This simplification also has the advantage of enabling the package to have greater computational efficiency when fitting higher-dimensional models. The above-mentioned recommendation to switch from an unstructured to an independent random effect covariance matrix at the 10 random effect predictor mark is an ad hoc recommendation determined by our experience creating and testing this package.

The MCECM algorithm outlined in Algorithm \ref{alg:MCECM} describes how the \CRANpkg{glmmPen}
package estimates the model parameters for a single set of penalty
parameters \((\lambda_0, \lambda_1)\). Section ``Tuning parameter selection''
discusses how the package chooses optimal set of tuning parameters during the model
selection procedure.

\section{Tuning parameter selection}
\label{sec:model-selection}


This section provides details on how the \code{glmmPen} function selects the set of optimal tuning parameters from a prespecific grid of values. Section ``Software'' provides further details on how to use both the \code{glmmPen} and \code{glmm} functions, where the latter function allows the user to fit a single model without performing variable selection on the fixed and random effects. 

For \code{glmmPen}, we generally recommend
that the user specify the `full model', i.e.,~specify the set of candidate random effects predictors to be equal to the set of candidate fixed effects predictors,
and let the algorithm select the best fixed and random effects using the procedure outlined in this section. However, if the user has some prior knowledge about the form of the random effects, they can restrict the random effects considered to an appropriate subset. As discussed in the previous section, the package requires that the random effects be a subset of the fixed effects. 

\subsection{Penalty sequence specification}
The \CRANpkg{glmmPen} package calculates default sequences of
penalty values for \(\lambda_0\) (penalizing the fixed effects
\(\boldsymbol \beta\)) and \(\lambda_1\) (penalizing the random effects
\(\boldsymbol \gamma\)), but allows users to enter their own
penalty sequences if desired.  We define the penalty parameter sequences for the fixed and random effects as
\(\boldsymbol \lambda_0 = (\lambda_{0,1},...,\lambda_{0,\omega_0})\) and
\(\boldsymbol \lambda_1 = (\lambda_{1,1},...,\lambda_{1,\omega_1})\), respectively, where $\omega_0$ and $\omega_1$ are the length of the fixed and random effect penalty sequences. These sequences are ordered from the minimum penalty ($\lambda_{0,1} = \lambda_{0,min}$ and $\lambda_{1,1} = \lambda_{1,min}$) to the maximum penalty ($\lambda_{0,\omega_0} = \lambda_{0,max}$ and $\lambda_{1,\omega_1} = \lambda_{1,max}$). By default, these sequences are calculated in a similar manner to the approach used by the package \CRANpkg{ncvreg} \citep{ncvreg2011}. The maximum penalty parameter
\(\lambda_{max}\) is calculated using the same procedure as \CRANpkg{ncvreg}; this value is assumed to penalize all fixed and random effects
coefficients to 0. We then set the sequence of penalty
parameters \(\boldsymbol \lambda_0 = \boldsymbol \lambda_1\) such that $\lambda_{0,max} = \lambda_{1,max} = 
\lambda_{max}$ and $\lambda_{0,min} = \lambda_{1,min} = \lambda_{min}$, where the minimum penalty parameter \(\lambda_{min}\) is a
small portion of the \(\lambda_{max}\). More details about these default sequences are given in Section ``Software''. In Section ``Tuning parameter selection strategy'', we consider a generic case where the $\boldsymbol \lambda_0$ and $\boldsymbol \lambda_1$ sequences do not need to be equal.

\subsection{Tuning parameter selection strategy}
\label{sec:search}
By default, the algorithm runs a computationally efficient two-stage approach to pick the optimal set of tuning parameters. In the first stage, the algorithm fits a
sequence of models where the fixed effect penalty is kept constant at
the minimum value of \(\boldsymbol \lambda_0\), \(\lambda_{0,min}\), and
the random effects penalty proceeds from the minimum value of
\(\boldsymbol \lambda_1\), \(\lambda_{1,min}\), to the maximum value
\(\lambda_{1,max}\). The optimal tuning parameter from this first stage is then
identified using Bayesian information criterion (BIC) type selection criteria, described in more detail
later in this section. This first stage identifies the optimal random
effect penalty value, \(\lambda_{1,opt}\). In the second stage, the
algorithm fits a sequence of models where the random effects penalty is
kept fixed at \(\lambda_{1,opt}\) and the fixed effects penalty
\(\boldsymbol \lambda_0\) proceeds from \(\lambda_{0,min}\) to
\(\lambda_{0,max}\). The overall best model is chosen from the models in
the second stage. In both stages, the results from each model are used
to initialize the coefficients in the subsequent model in the sequence.

Unlike other packages that perform variable selection, such as \CRANpkg{ncvreg}
and \CRANpkg{grpreg}, we run the \(\boldsymbol \lambda_0\) and $\boldsymbol \lambda_1$ sequences from their minimum value to their maximum value and not the traditional progression from their maximum value to their minimum value. In this mixed model setting, we have found (through simulations conducted during early package testing) that
this approach provides better initialization of subsequent models in the
tuning parameter sequence, giving an overall better performance to the algorithm
and improving algorithm speed. 
This progression of penalty sequences also speeds up the overall variable selection procedure by restricting the random effects considered during later penalty combinations within the variable selection procedure.
Within stage one, if a previous tuning parameter in the grid penalized out
a set of random effects from the model, the following model in the tuning parameter sequence will automatically
ignore these random effects. Within stage two, the random effects considered
are restricted to the non-zero random effects from the best model in
stage one.

In the original MCECM algorithm implementation given in \cite{rashid2020}, the authors searched for the best model by performing a `full grid search' and evaluating all possible combinations of $\boldsymbol \lambda_0$ and $\boldsymbol \lambda_1$. (We sometimes refer to the two-stage approach as the `abbreviated grid search'). While the \CRANpkg{glmmPen} package can perform this full grid search, we strongly recommend the two-stage abbreviated grid search. Compared with the full grid search, the two-stage grid search significantly reduces the required time to complete the algorithm, particularly when the number of random effects predictors is large. Furthermore, we have found that the two-stage grid search works very well in practice (see Section
``Simulations'' for performance results).

If users wish to perform a full grid search, the path of solutions is initialized by fitting a model using the minimum penalty for both the fixed and random effects
(\(\lambda_{0,1} = \lambda_{0,min}, \lambda_{1,1} = \lambda_{1,min}\)).
The algorithm then proceeds to estimate models over the full grid of
\(\boldsymbol \lambda_0\) and \(\boldsymbol \lambda_1\). For each value
of \(\lambda_{1,h} \in \boldsymbol \lambda_1\) that penalizes the random
effects, the fixed effects penalty parameter sequence proceeds from the
minimum value \(\lambda_{0,min}\) to the maximum value
\(\lambda_{0,max}\) while keeping \(\lambda_{1,h}\) fixed. Each model is initialized using
the result from the model fit with the previous tuning parameter combination in the sequence. The algorithm then updates the
penalty parameter to the next \(\lambda_{1,h+1}\) and repeats the
process. The model with the penalty parameter combination (\(\lambda_{0,min}\),\(\lambda_{1,h+1}\)) is initialized using the model from the previous ($\lambda_{0,min},\lambda_{1,h}$) penalty parameter combination.

\subsection{Optimal tuning parameter selection}
\label{sec:BIC-discussion}
Once models have been fit pertaining to all tuning parameter combinations
within the first and second stages of the tuning parameter search strategy (or over the full tuning parameter grid search), the \CRANpkg{glmmPen} package chooses the best model from one of several BIC-type selection criteria options. For simplification of notation, consider the generic penalty parameter combination $\lambda = (\lambda_0,\lambda_1)$ that penalizes the fixed and random effects, respectively. By default, the package uses the BIC-ICQ criterion \citep{BICq2011}, where the abbreviation ICQ stands for ``Information Criterion based on the Q-function''. This BIC-ICQ criteria is expressed below:
\begin{equation}
    \begin{aligned}
    \text{BICq}(\boldsymbol \theta_\lambda) &= 2 \{ Q_1(\boldsymbol \theta_\lambda | \boldsymbol \alpha_0) + Q_2(\boldsymbol \alpha_0)\} + d_\lambda  \log(N) \\
    &\approx \left \{-\frac{2}{M} \sum_{m=1}^M \sum_{k=1}^K \left [ \log f(\boldsymbol y_k | \boldsymbol X_k, \boldsymbol \alpha_{0,k}^{(m)}; \boldsymbol \theta_\lambda) + \log \phi (\alpha_{0,k}^{(m)}) \right ] \right \}  
    + d_\lambda  \log(N),
\end{aligned}
\end{equation}
where $\boldsymbol \theta_\lambda$ are the coefficients of the model fit with the penalty $\lambda = (\lambda_0, \lambda_1)$, $\boldsymbol \alpha_0$ are the posterior samples from a ``minimal penalty model''---the model with either no penalty (when the number of random effects predictors is less than 5) or a minimum penalty used on the fixed and random effects---and $\alpha_{0,k}^{(m)}$ is the $m^{th}$ posterior sample for group $k$ from such a minimal penalty model, $Q_1$ and $Q_2$ were defined in Section ``MCECM algorithm'', \(d_\lambda\) is the number of nonzero coefficients for the model (all nonzero \(\boldsymbol \beta\) plus all nonzero \(\boldsymbol \gamma\)), and \(N\) is the total number of observations in the data (\(N_{obs}\)).


The package can also calculate the traditional BIC criterion as specified below: \begin{equation*}
  \text{BIC}(\boldsymbol\theta_\lambda) = -2\ell(\boldsymbol\theta_\lambda) + d_\lambda \log(N),
\end{equation*} where \(\boldsymbol \theta_\lambda\) are the
coefficients of the penalization model,
\(\ell(\boldsymbol \theta_\lambda)\) is the marginal log-likelihood for
the model, \(d_\lambda\) is the number of nonzero coefficients for the
model, and \(N\) can be either the total number of observations in the data (\(N_{obs}\)) or the total number of independent observations (i.e.,~number of levels within the grouping factor, \(N_{grps}\)) in the data. The marginal log-likelihood is as
follows: \begin{equation}
  \ell(\boldsymbol \theta) = \sum_{k=1}^K \ell_k(\boldsymbol \theta) = \sum_{k=1}^K \frac{1}{n_k} \log \int f(\boldsymbol y_k | \boldsymbol X_k, \boldsymbol \alpha_k; \boldsymbol \theta) \phi(\boldsymbol \alpha_k) d \boldsymbol \alpha_k.
\end{equation}

There is a lack of consensus regarding the use of $\log(N_{obs})$ versus $\log(N_{grps})$ in the BIC penalty term for mixed models. For instance, the $\log(N_{obs})$ penalty is
used in the R package \CRANpkg{nlme} \citep{nlmeManual}, and the $\log(N_{grp})$ penalty is used in SAS proc NLMIXED \citep{SAS,BICh2014}. In practice, the performance of the different versions of the BIC penalty term may depend on the true underlying model \citep{lorah2019value,BICh2014}, with Delattre et al. \citeyearpar{BICh2014} observing that the $\log(N_{obs})$ penalty performed better when the true model had very few random components, and the $\log(N_{grp})$ penalty performed better when the true model had a large number of random components. Both Delattre et al. \citeyearpar{BICh2014} and Lorah and Womack \citeyearpar{lorah2019value} suggest using some combination of these sample size definitions.

To this point, the package also allows the best model to be selected using a `hybrid' BICh selection criteria developed by Delattre et al. \citeyearpar{BICh2014}: \begin{equation}
  \text{BICh}(\boldsymbol\theta_\lambda) = -2\ell(\theta_\lambda) + d_{\lambda,\beta} \log(N_{obs}) + d_{\lambda,\gamma} \log(N_{grps}),
\end{equation} where \(d_{\lambda,\beta}\) and \(d_{\lambda,\gamma}\) are the number of nonzero fixed and random effect coefficients, respectively.


In simulations not shown here (see content in GitHub repository \url{https://github.com/hheiling/paper_glmmPen_RJournal} for details), we found that the BIC-ICQ gave the best performance in choosing the correct set of fixed and random effects. The BIC and BICh methods tended to underestimate the number of true fixed effects compared to BIC-ICQ in the simulations we considered. However, in order to calculate the BIC-ICQ, a minimal penalty model needs to be fit using a small penalty (i.e.,~$\lambda_{min}$) on the fixed and random effects. Posterior samples from this minimal penalty model are then used to calculate the BIC-ICQ value for each model fit in the variable selection procedure. Depending on the size of the full model with all fixed and random effects predictors, this calculation can be time-intensive since fitting the model with a small penalty will keep many fixed and random effects predictors in the model. 

Alternatively, the calculation of the BIC and BICh criteria require a calculation of the marginal log-likelihood \(\ell(\boldsymbol \theta)\) for each model. Since the integrals within \(\ell(\boldsymbol \theta)\) are intractable, we estimate the marginal log-likelihood using the corrected arithmetic mean estimator (CAME) described by Pajor \citeyearpar{pajor2017}. We have found this CAME estimator to be relatively quick and easy to calculate, as well as consistent with the marginal log-likelihood estimate calculated by the package \CRANpkg{lme4} \citep{lme42007} for a range of conditions (see content in GitHub repository \url{https://github.com/hheiling/paper_glmmPen_RJournal} for details).

To calculate the CAME, we focus on a single group \(k\) and define a set
\(A_k \subseteq \Theta\) as a subset of the parameter space of the
random effects for group \(k\), where \(P(A_k)\) and
\(P(A_k |\boldsymbol y_k, \boldsymbol X_k; \boldsymbol \theta)\) are
nonzero probabilities. We first start with the knowledge \begin{align}
\begin{aligned}
  P(A_k | \boldsymbol y_k, \boldsymbol X_k; \boldsymbol\theta) & = \int_{A_k} \phi(\boldsymbol\alpha_k | \boldsymbol y_k, \boldsymbol X_k; \boldsymbol\theta) d \boldsymbol\alpha_k \\
  & = \int_\Theta \frac{1}{f(\boldsymbol y_k | \boldsymbol X_k; \boldsymbol\theta)} f(\boldsymbol y_k | \boldsymbol X_k, \boldsymbol\alpha_k; \boldsymbol\theta) \phi(\boldsymbol\alpha_k) I(\boldsymbol\alpha_k \in A_k) d \boldsymbol\alpha_k,
\end{aligned}
\end{align} where I(.) is an indicator function,
\(f(\boldsymbol y_k | \boldsymbol X_k; \boldsymbol \theta) = \int f(\boldsymbol y_k | \boldsymbol X_k, \boldsymbol \alpha_k; \boldsymbol \theta) \phi(\boldsymbol \alpha_k) d \boldsymbol \alpha_k\)
is the marginal likelihood for group \(k\), and all other terms are
described in Section ``Generalized linear mixed models''. The above relationship allows
us to obtain the result: \begin{align}
\begin{aligned}
  f(\boldsymbol y_k | \boldsymbol X_k; \boldsymbol\theta) & = \frac{1}{P(A_k|\boldsymbol y_k, \boldsymbol X_k; \boldsymbol\theta)} \int_\Theta f(\boldsymbol y_k | \boldsymbol X_k, \boldsymbol\alpha_k; \boldsymbol\theta) \phi(\boldsymbol\alpha_k) I(\boldsymbol\alpha_k \in A_k) d \boldsymbol\alpha_k \\
  & = \frac{1}{P(A_k|\boldsymbol y_k, \boldsymbol X_k; \boldsymbol\theta)} \int_\Theta \frac{f(\boldsymbol y_k | \boldsymbol X_k, \boldsymbol\alpha_k; \boldsymbol\theta) \phi(\boldsymbol\alpha_k) I(\boldsymbol\alpha_k \in A_k) s(\boldsymbol\alpha_k) d \boldsymbol\alpha_k}{s(\boldsymbol\alpha_k)},
  \label{eqn:pajor}
\end{aligned}
\end{align} where \(s(.)\) is an importance sampling function.

Suppose at the end of the MCECM algorithm we obtain \(M\) samples from
the posterior distribution of the random effects for group \(k\),
\(\boldsymbol{\tilde \alpha}_k= ((\boldsymbol \alpha_k^{(1)})^\top,...,(\boldsymbol \alpha_k^{(M)})^\top)^\top\).
Let us set \(A_k = \boldsymbol{\tilde \alpha}_k\); this reduces
\(P(A_k | \boldsymbol y_k, \boldsymbol X_k; \boldsymbol \theta)\) to 1.
Let us also set the importance sampling function \(s(.)\) to be a
multivariate normal distribution with a mean vector equal to the mean of
the posterior samples
\(\frac{1}{M}\sum_{m=1}^M \boldsymbol \alpha_k^{(m)}\) and a covariance
matrix equal to the covariance matrix of a thinned subset of the
posterior samples (to obtain a pseudo-independent set of samples). If we draw \(M^\star\) samples
\(\boldsymbol\alpha_k^\star= ((\boldsymbol \alpha_k^{\star(1)})^\top,...,(\boldsymbol \alpha_k^{\star(M^\star)})^\top)^\top\)
from this importance sampling function, then Equation~\ref{eqn:pajor}
indicates that we can estimate the marginal likelihood for group \(k\)
as \begin{align}
\begin{aligned}
  f(\boldsymbol y_k | \boldsymbol X_k; \boldsymbol\theta) \approx \frac{1}{M^\star} \sum_{m=1}^{M^\star} 
  \frac{f(\boldsymbol y_k | \boldsymbol X_k, \boldsymbol\alpha_k^{\star m}; \boldsymbol\theta) \phi(\boldsymbol\alpha_k^{\star m}) I(\boldsymbol\alpha_k^{\star m} \in A_k)}{s(\boldsymbol\alpha_k^{\star m})}.
  \label{eqn:pajor_est}
\end{aligned}
\end{align}

We repeat the estimation in Equation~\ref{eqn:pajor_est} for all \(K\) groups in order to calculate the full desired marginal log-likelihood \(\ell(\boldsymbol \theta)\). This final marginal log-likelihood is then used in the previously mentioned BIC and BICh calculations for each fitted model across the \(\boldsymbol \lambda_0\) and \(\boldsymbol \lambda_1\) grid search. We refer to this marginal log-likelihood as the Pajor log-likelihood in later sections of the paper.

\section{Software}
\label{sec:software}

The main function of the \CRANpkg{glmmPen} package is \code{glmmPen}, which is used to perform fixed and random effects variable selection after the specification of a full model with all candidate fixed and random effects. The \CRANpkg{glmmPen} package is also capable of fitting a GLMM with pre-specified fixed and random effects (under no penalization) using the function  \code{glmm}. Here we will use the \code{basal} dataset \citep{rashid2020} to illustrate the use of the \code{glmmPen} function in practical applications. 



\subsection{Data example}
\label{sec:examples}

The \code{basal} dataset is composed of four studies that contain gene expression data and tumor subtype information from patients spanning three cancer types \citep{moffitt2015,weinstein2013}. Two of these datasets contain
gene expression data for subjects with Pancreatic Ductal Adenocarcinoma
(PDAC), one dataset contains data for subjects with Breast Cancer, and
the last contains data for subjects with Bladder Cancer. While each cancer type has separate sets of defined subtypes, all share a common subtype defined as ``basal-like'', which was shown to be similar in character across cancer types and have an impact on survival \citep{moffitt2015}.  The goal of the original study was to select features that are relevant in predicting the basal-like subtype.  To increase the sample size, it was proposed that samples were merged from each study into one large dataset. 

Multiple approaches have been proposed to integrate gene expression data from multiple studies to improve the accuracy of downstream prediction models \citep{riester2014, ma2018continuity, patil2018}.  The pGLMM methodology from \cite{rashid2020} was originally motivated by the need to select genes that are predictive of cancer outcomes (e.g. cancer subtype), where the effects of genes may vary randomly across studies.  It was shown that accounting for this heterogeneity improved the performance of gene selection after data merging.

Using \CRANpkg{glmmPen} package, we fit a pGLMM that can accommodate a large number of features in the model and account the hetereogeneity in gene effects across studies.  It is unclear \textit{a priori} which features truly have a non-zero association with the outcome and which features truly have variation in their effects across studies. Therefore, we will use the \code{glmmPen} function to simultaneously select fixed and random effects from a set of candidate features. \cite{rashid2020} integrated gene expression data from each study using a binary rank transformation technique (Top Scoring Pairs or TSPs), which we use as our covariates in this example. To illustrate the concept of TSPs, let $g_{ki,A}$ and $g_{ki,B}$ be the raw expression of genes $A$ and $B$ in subject $i$ of group $k$. For each gene pair ($g_{ki,A}$, $g_{ki,B}$), the TSP is the indicator $I(g_{ki,A} > g_{ki,B})$ which specifies which gene of the two has higher expression in the subject. We denote a TSP predictor as ``GeneA\_GeneB''.  A total of 50 binary TSP covariates are provided in the \code{basal} dataset available in the package. For illustration purposes, we randomly select 10 TSP covariates. Our goal is to identify TSPs that are associated with patient tumor subtype while accounting for study-level heterogeneity in gene effects.  In each study subtype is defined a binary variable with two levels: basal-like or non-basal-like.  Therefore, for this example, our example dataset consists of our matrix of covariates $X$, our subtype vector $y$ (a factor with two levels), and our study membership vector (a factor with four levels). 

\newpage
Summary information about the data is included below.

\begin{example}
> library("glmmPen")
> data("basal")
> y = basal$y
> set.seed(1618)
> idx = sample(1:50, size = 10, replace = FALSE)
> idx = idx[order(idx)]
> X = basal$X[,idx]
> colnames(X)
\end{example}
\begin{example}
[1] "GPR160_CD109"   "SPDEF_MFI2"    "CHST6_CAPN9"   "SLC40A1_CDH3"  
[5] "PLEK2_HSD17B2"  "GPX2_ERO1L"    "CYP3A5_B3GNT5" "LY6D_ATP2C2"   
[9] "MYO1A_FGFBP1"   "CTSE_COL17A1"  
\end{example}

\begin{example}
> group = basal$group
> levels(group)
\end{example}
\begin{example}
[1] "UNC_PDAC"     "TCGA_PDAC"    "TCGA_Bladder" "UNC_Breast"  
\end{example}


We will fit a penalized random effects logistic regression model using the \code{glmmPen} function to model patient subtype, as it is unclear which of the 10 TSPs should be included in the model, and which may also randomly vary across studies in their effects.  We perform variable selection using the following code:

\begin{example}
> set.seed(1618)
> fitB = glmmPen(formula = y ~ X + (X | group), 
+                family = "binomial", covar = "independent",
+                tuning_options = selectControl(BIC_option = "BICq", 
+                                               pre_screen = TRUE, 
+                                               search = "abbrev"),
+                penalty = "MCP", BICq_posterior = "Basal_Posterior_Draws")
\end{example}

Here we utilize the pre-screening and abbreviated grid search options, as well as select the optimal tuning parameter using the BIC-ICQ model selection criteria (denoted ``BICq''). Further details about the pre-screening procedure is described in the Section ``selectControl arugments'' and the consequences of this pre-screening procedure are illustrated through simulations and discussed in Section ``Pre-screening performance''. If we were instead interested in fitting a GLMM utilizing all 10 TSPs as fixed effects and assuming a random effect for each (without penalization), we could run the following code:

\begin{example}
> set.seed(1618)
> fit_glmm = glmm(formula = y ~ X + (X | group), 
+                  family = "binomial", covar = "independent", 
+                  optim_options = optimControl())
\end{example}

The set of random effects specified does not necessarily have to be equal to the set of fixed effects as in the above example. Because of the number of random effects that we are considering in the model, we approximate the random effects covariance matrix as an independent, or diagonal, matrix, which we specify by using the argument option \code{covar = "independent"}. Our reasoning for such an approximation, as well as a discussion of the pros and cons of such an approximation, are given in Section ``Initialization and convergence.''

In the following subsections, we will discuss in detail the \code{glmmPen} (and \code{glmm}) arguments and relevant output. We will also examine the output from the variable selection procedure given by the \code{glmmPen} example.

\subsection{Full model specification}
\label{sec:model-specification}

The syntax for specifying the full model formula (the model with all relevant fixed and random effects predictors) using the \code{formula} argument closely follows the formula syntax of
the \CRANpkg{lme4} package \citep{lme42007}. The formula follows the form
\code{response\ \textasciitilde{}\ fix\_expr\ +\ (rand\_expr\ \textbar{}\ factor)}
where the \code{fix\_expr} specifies the variables to use as the fixed
effects, the \code{rand\_expr} specifies the variables to use as the
random effects, and the \code{factor} specifies the grouping factor of
the observations. When a data frame is given for the \code{data}
argument, the fixed and random effects can be specified using the column
names of the data frame. For higher-dimensional data, users
may find it easier to directly specify the matrix containing the covariates of interest and the response vector, such as the
\code{y\ \textasciitilde{}\ X\ +\ (X\ \textbar{}\ group)} formula
given in the earlier \code{glmmPen} fit example. No specification of
the \code{data} argument is needed in this case. Similar to \CRANpkg{ncvreg}, an intercept is always assumed and required, and therefore an intercept column need not be specified in $X$ or explicitly in the model formula; \texttt{glmmPen} will output an error if the input predictor matrix $X$ contains an intercept column. 

Regarding the specification of random effects in \code{formula}, the \CRANpkg{glmmPen} package currently does not allow for multiple grouping factors.  In addition, the random effects must be a subset of the fixed effects, and a random intercept is always assumed and required in the model. Lastly, the structure of the random effects covariance matrix is determined by the \code{covar} argument, which may take on the value of  `unstructured' or `independent' (diagonal). By default, the \code{covar} parameter is set to \code{NULL}. This automatically selects the `independent' option if the number of random effect predictors is 10 or more and selects  `unstructured' otherwise. For a large number of random effect predictors, it is strongly recommend that the covariance structure to `independent' in order to improve computational efficiency.

The \code{glmmPen} algorithm allows the Binomial, Gaussian, and
Poisson families with canonical links.

\subsection{Penalization and optimal tuning parameter selection}
\label{sec:selection-penalization}

In \code{glmm}, the default is to fit the single model with user-specified fixed and random effects with no penalization. Although it is generally not recommended, users have the option to specify a single penalty parameter combination using \\
\code{tuning\_options = lambdaControl(lambda0,lambda1)}. In \code{glmmPen}, the arguments \code{penalty}, \\
\code{gamma\_penalty}, \code{alpha}, \code{fixef\_noPen}, and \code{tuning\_options} all
play a part in the variable selection process. The following subsections
discuss these argument options in detail and how the arguments impact
variable selection.

\subsubsection{Penalty, gamma penalty, alpha
parameters}
\label{sec:penalty-params}

To perform variable selection, \code{glmmPen} allows the fixed effect coefficients to
be penalized using the minimax concave penalty (MCP, the default), the smoothly clipped absolute deviation (SCAD) penalty, or the least absolute shrinkage and selection operator (LASSO) penalty \citep{ncvreg2011, glmnet2010} via the \code{penalty} argument, which takes as input the character strings ``MCP'', ``SCAD'', or ``lasso''. The random effects are then penalized using the grouped version of the selected penalty type \citep{grpreg2015}, e.g.,~if the MCP penalty is used to penalize the fixed effects, then the grouped MCP penalty is used to penalize the random effects covariance matrix coefficients. 

In addition to the previously discussed penalty parameters ($\lambda_0, \lambda_1$), the MCP and SCAD penalties also use a scaling factor \citep{ncvreg2011, grpreg2015}. The argument \code{gamma\_penalty}
specifies this scaling factor, with the default of 3 and 4 for the MCP and SCAD penalties, respectively. Additionally, the argument
\code{alpha} allows for the elastic net estimator, controlling the relative
contribution of the MCP/SCAD/LASSO penalty and the ridge, or $L_2$,
penalty. Setting \code{alpha} to 1 (the default) is equivalent to the
regular penalty with no $L_2$ contribution.

\subsubsection{selectControl arguments}
\label{sec:selectControl}

The grid search over the fixed effects and random effects penalty parameters \(\boldsymbol \lambda_0\) and \(\boldsymbol \lambda_1\) is controlled by the arguments in \code{selectControl()}. The user can specify particular sequences for $\boldsymbol \lambda_0$ (fixed effects penalty parameters) and $\boldsymbol \lambda_1$ (random effects penalty parameters) using the arguments \code{lambda0\_seq} and \code{lambda1\_seq}, respectively; by default, a sequence of penalty parameters (of length \code{nlambda}, default 10) are automatically calculated within \code{glmmPen}. These default sequences are calculated using the method discussed in Section ``Tuning parameter selection''.
The minimum penalty \(\lambda_{min}\) is a small fraction
of the \(\lambda_{max}\) value; the argument \code{lambda.min} controls
what fraction is used. By default,  \code{lambda.min} = 0.01 so that
\(\lambda_{min} = 0.01(\lambda_{max})\).

The structure of the optimal tuning parameter search is specified by the argument
\code{search}. If \code{search} = ``abbrev'' (default), the
algorithm performs the abbreviated two-stage tuning parameter search specified in
Section ``Tuning parameter selection''. If \code{search} = ``full\_grid'', the
algorithm looks over the full grid search of length(\code{lambda0\_seq})$\times$length(\code{lambda1\_seq}) models
before picking the best model.

After all of the tuning parameters have been evaluated, the optimal combination of tuning parameters can be selected using a BIC-type selection criteria, which can be specified
using the \code{selectControl()} argument \code{BIC\_option}. Using
the \code{BIC\_option} argument, the user can select one of four
BIC-type selection criteria, given in Table \ref{tab:BIC_options}, to
select the best model.

\begin{table}[h!]
  
  \centering
  \begin{tabular}{cp{4in}}
  \toprule
  Selection criteria & Description \\
  \midrule
  BICq & (Default) BIC-ICQ selection criteria (Ibrahim et al., 2011); requires fitting the minimal penalty model \\
  BICh & Alternative BICh selection criteria specified by Delattre, Lavielle, and Poursat (2014) \\
  BIC & Traditional BIC whose penalty term sets $N$ to the number of total observations in the data \\
  BICNgrp & Traditional BIC whose penalty term sets $N$ to the number of independent observations (i.e., number of levels of the grouping factor) \\
  \bottomrule
  \end{tabular}
  \caption{BIC-type model selection criteria options for argument BIC\_option.}
  \label{tab:BIC_options}
\end{table}

Refer to the discussion in Section ``Tuning parameter selection'' for further
details about these BIC-type options as well as their respective pros
and cons. 


The argument \code{pre\_screen} allows users to screen out some random
effects at the start of the algorithm. When \code{pre\_screen} is set
to \code{TRUE} (the default) and the number of random effects predictors is 5 or more, a minimal penalty model is fit using a small penalty
for the fixed and random effects and relatively lax convergence
criteria. If at the end of the pre-screening procedure the variance of a
random effect is penalized to 0 or is estimated to be less than
\(10^{-2}\), that predictor is restricted to have a zero-valued random
effect variance for all models fit by the algorithm. The pre-screening
procedure is not implemented if the number of random effects is less
than five. This threshold of five random effect predictors is an ad hoc choice by the authors; the purpose of the pre-screening procedure is to allow the user to speed up the variable selection procedure when the full model contains a large number of random effects.

The argument \code{lambda.min.presc} adjusts the value of the random effect penalty parameter $\lambda_1$ used in the pre-screening step and the minimal penalty model fit for the BIC-ICQ calculation, where the minimum penalty used for the random effects is \code{lambda.min.presc} \(\times \lambda_{max}\). See package documentation for further details about this argument and other minor arguments not discussed here.

\subsubsection{Additional selection arguments in
glmmPen}
\label{sec:additional}

The default variable selection procedure assumes that we have no prior
knowledge of which fixed effects should not be penalized during the model fitting procedure. In order to indicate that a covariate should not be subject to penalization (and therefore always remain in the model), one can use the \code{fixef\_noPen} argument. See the \code{glmmPen} function documentation for further details. 


After running an initial grid search over the default fixed and random effects penalty parameters, users may desire to re-run the variable selection procedure using alternative settings, such as different penalty sequences (e.g.~a finer grid search) or different convergence criteria. In this scenario, re-computing the the minimal penalty model for the BIC-ICQ criterion calculation can be time-consuming. In order to save the minimal penalty model posterior samples
needed for the BIC-ICQ calculation and re-use these samples to compute the BIC-ICQ within a subsequent tuning parameter selection grid search, the user can save the posterior samples as a file-backed \code{big.matrix} using the argument \code{BICq\_posterior} = ``file\_location/file\_name''.  This saves the backing file and the descriptor file as \file{file\_location/file\_name.bin} and \file{file\_location/file\_name.desc}, respectively. If the
file name is not specified, then the posterior samples are
automatically saved to the working directory with the file name ``BICq\_Posterior\_Draws''. These saved posterior samples can then be re-loaded in R as a \code{big.matrix} using the \code{attach.big.matrix} function from the package
\CRANpkg{bigmemory} \citep{bigmemory2013}. In secondary calculations using \code{glmmPen}, the recalculation of this minimal penalty model fit can be avoided and these posterior samples can be used by calling
\code{BICq\_posterior} = ``file\_location/file\_name''.

\subsection{Examination of output}
\label{sec:exam-output}

The \code{glmm} and \code{glmmPen} functions all output a reference class object of class \code{pglmmObj}. A full list of the methods available for \code{pglmmObj} objects are provided in Table \ref{tab:methods}. These methods and their output were designed to be very similar to the methods and output available for \code{merMod} objects used in the \CRANpkg{lme4} package. Further information about the output provided in a \code{pglmmObj} object and additional methods documentation is available in the \CRANpkg{glmmPen} package documentation (see \code{?pglmmObj}).

\begin{table}[h!]
  
  \centering
  \begin{tabular}{lp{5in}}
  \toprule
  Generic & Brief description of return value \\
  \midrule
  BIC & Numeric vector returning the BIC, BICh, BICNgrp, and, if specified for model selection, BIC-ICQ selection criteria evaluations for either the
  fitted \code{glmm} model or the optimal fitted \code{glmmPen} model (i.e. the `best' model according to the model selection criteria) \\
  coef & Matrix reporting the sum of the fixed effects coefficients and the posterior modes of the random effects for each variable at each level of the grouping factor \\
  fitted & Numeric vector of fitted values (the values of the linear predictor) based on either the fixed effects only (recommended for most applications) or both the fixed effects and the posterior modes of the random effects for each level of the grouping factor (potentially useful for diagnostics) \\
  fixef & Numeric vector of the fixed effects coefficient estimates $\boldsymbol{\widehat{\beta}}$\\
  formula & The mixed-model formula of the fitted model \\
  logLik & Estimated log-likelihood for the best model of the \code{glmmPen} procedure or the final model from \code{glmm} evaluated using the Pajor (2017) marginal likelihood calculation discussed in Section ``Tuning parameter selection'' \\
  model.frame & A data.frame object containing the output and predictors used to fit the model\\
  model.matrix & The fixed-effects model matrix \\
  ngrps & Number of levels in the grouping factor \\
  nobs & Number of total observations \\
  plot & Diagnostic plots for mixed-model fits \\
  predict & Predicted values based on either the fixed effects only (recommended) or the combined fixed effects and posterior modes of the random effects for each variable and each level of the grouping factor \\
  print & Basic printout of mixed-model objects \\
  ranef & Matrix of posterior modes of the random effects for each variable and each level of the grouping factor \\
  residuals & Numeric vector of residual values: deviance (default), Pearson, response, or working residuals \\
  sigma & Random effect covariance matrix ($\boldsymbol{\Gamma\Gamma}^\top$) \\
  summary & Summary of the mixed model results \\
  \bottomrule
  \end{tabular}
  \caption{List of currently available methods for objects of class pglmmObj.}
  \label{tab:methods}
\end{table}

When the \code{pglmmObj} object is created using the \code{glmm} function, the output from the methods listed in Table \ref{tab:methods} pertains to the single model fit specified by the \code{glmm} arguments. When the \code{pglmmObj} object is created using the \code{glmmPen} function, the output from the methods pertains to the best model chosen during the model
selection procedure. Additional information about each model fit can be found in the \code{results\_all} field of the \code{pglmmObj} object. Using the \code{basal} output object
\code{fitB} from the \code{glmmPen} function, we illustrate the use of several of these methods in the remainder of this section.

\subsubsection{Model summary}
\label{sec:model-summary}

The \code{summary} method output the function call information such as the sampler used in the E-step (in this case, \pkg{Stan}), the family, the model formula, the estimates of the fixed effects, the variance and standard deviation estimates of the random effects, and a summary of the deviance residuals. (Note: Due to the style of our formula specification using a matrix instead of column names of a data.frame, all variable names begin with the name of the matrix, \code{X}.)

\begin{example}
> summary(fitB)
\end{example}
\begin{example}
 Penalized generalized linear mixed model fit by Monte Carlo Expectation 
  Conditional Minimization (MCECM) algorithm (Stan)  ['pglmmObj'] 
  Family: binomial  ( logit )
 Formula: y ~ X + (X | group)
 
 Fixed Effects:
   (Intercept)   XGPR160_CD109     XSPDEF_MFI2    XCHST6_CAPN9   XSLC40A1_CDH3  
       -1.1530         -0.7099         -0.7355          0.5082         -0.5831 
 XPLEK2_HSD17B2     XGPX2_ERO1L  XCYP3A5_B3GNT5    XLY6D_ATP2C2   XMYO1A_FGFBP1  
        0.4337         -0.5895          0.0000          0.4620         -0.7411 
  XCTSE_COL17A1  
        0.0000  

Random Effects:
 Group Name           Variance Std.Dev.
 group (Intercept)    0.8193   0.9052  
 group XGPR160_CD109  0.2036   0.4512  
 group XSPDEF_MFI2    0.684    0.827   
 group XCHST6_CAPN9   0        0       
 group XSLC40A1_CDH3  0        0       
 group XPLEK2_HSD17B2 0.0804   0.2835  
 group XGPX2_ERO1L    0.0842   0.2901  
 group XCYP3A5_B3GNT5 0        0       
 group XLY6D_ATP2C2   0        0       
 group XMYO1A_FGFBP1  0.1551   0.3938  
 group XCTSE_COL17A1  0.7596   0.8715   
Number Observations: 938,  groups: group, 4 

Deviance residuals:  
    Min      1Q  Median      3Q     Max 
-2.9338 -0.4026 -0.1512  0.3457  2.9630
\end{example}

 



We see that the best model included 9 TSPs with non-zero fixed effects and 6 TSPs with non-zero random effects (i.e.,~6 TSPs with varying predictor effects across the studies). The \code{print} method supplies very similar information to the \code{summary} method minus the summary of the residuals.

The individual components of the \code{print} and \code{summary} outputs can be obtained using several accessory functions described in Table \ref{tab:methods}. Similar to the package \CRANpkg{lme4}, the fixed effects can be summarized using \code{fixef} and the group-specific random effects can be summarized using \code{ranef}. The random effect covariance matrix is summarized using \code{sigma}. In the case of the
Gaussian family, \code{sigma} also provides the residual standard error.


\begin{example}
> fixef(fitB)
> ranef(fitB)
> sigma(fitB)
\end{example}

The residuals for the final model can be called using the \code{residuals} method. The different \code{type} options for the residuals include
``deviance'', ``pearson'', ``response'', and ``working'', which
correspond to the deviance, Pearson, response, and working residuals, respectively.

\begin{example}
> residuals(fitB, type = "deviance")
\end{example}

\subsubsection{Predictions and fitted
values}
\label{sec:predictions-fitted}

Using the \code{predict} method, we can make predictions using only
the population level information (i.e.,~the fixed effects only) or the
group-specific level information (i.e.,~the fixed and random effects
results). The \CRANpkg{glmmPen}
package restricts predictions on new data to only use the fixed effects since it is generally unlikely that the grouping levels within other datasets will exactly match the grouping levels within the data used to create the prediction model.
The \code{predict} method has the following arguments:

\begin{itemize}
\item
  \code{object}: an object of class \code{pglmmObj} output from
  \code{glmm} or \code{glmmPen}.
\item
  \code{newdata}: a data frame of new data that contains all of the
  fixed effects covariates from the model fit. The variables provided in
  \code{newdata} must match the fixed effects used in the model fit.
\item
  \code{type}: a character string specifying whether to output the
  linear predictor (``link'', default) or the expected mean response
  (``response'').
\item
  \code{fixed.only}: boolean value specifying if the prediction is
  made with only the fixed effects (\code{TRUE}, default) or both the
  fixed and random effects (\code{FALSE}). Predictions are restricted
  to \code{fixed.only\ =\ TRUE} for new data predictions.
\end{itemize}

The \code{fitted} method also includes the \code{fixed.only}
argument, allowing the fitted values of the linear predictor to be
estimated with or without the random effects estimates.

\begin{example}
> predict(object = fitB, newdata = NULL, type = "link", fixed.only = TRUE)
> fitted(object = fitB, fixed.only = TRUE)
\end{example}

\subsubsection{Diagnostics}
\label{sec:diagnostics}

The \CRANpkg{glmmPen} package provides methods to perform diagnostics on the final model fit object. The \code{plot} method plots the residuals against the fitted values. The \code{plot} function defaults to plotting the Pearson residuals
for the Gaussian family, and deviance residuals otherwise.
\begin{example}
> plot(object = fitB)
\end{example}

The \code{plot\_mcmc} function performs graphical MCMC diagnostics on
the random effect posterior samples. 
This command has six arguments with
the first argument specifying the \code{pglmmObj} output object. The
second argument \code{plots} is used to specify which diagnostics
plots to produce. The \code{plots} argument is capable of
creating sample path plots (``sample.path'', default), autocorrelation plots
(``autocorr''), cumulative sum plots (``cumsum''), and histograms
(``histogram'') of the posterior samples. The plots are output as faceted
\CRANpkg{ggplot2} \citep{ggplot2book} plots with the graphics arranged by groups in the columns and
variables in the rows. As objects of class \code{ggplot}, they are capable of being edited as any other \code{ggplot} object. The \code{plots} argument can specify a vector
of multiple plot types or the choice of ``all'', which
automatically produces all four types of diagnostic plots. The function
outputs a list object containing the plots specified. The third and fourth arguments
\code{grps} and \code{vars} allow the user to restrict which groups
and/or variables are summarized in the diagnostic plots. The default
values of ``all'' for these arguments give the results for all groups
and variables. To request specific groups and variables, provide vectors
of character strings specifying the variable or group names. The argument
\code{numeric\_grp\_order} tells the function to order the group levels numerically (default \code{FALSE}), and \code{bin\_width} allows the
user to manipulate the bin widths of the histograms (default
\code{NULL} results in \code{geom\_histogram} defaults, only relevant if the ``histogram'' plot is requested).

\begin{figure}[h!]
    \centering
    \includegraphics[width=0.8\textwidth]{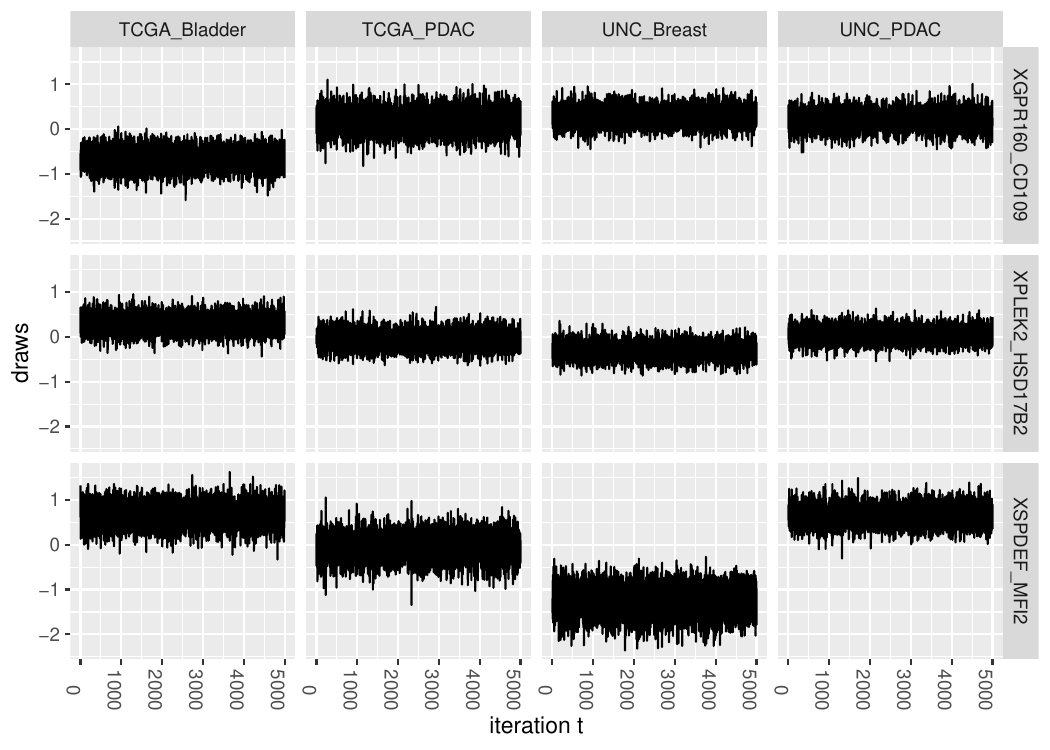}
    \includegraphics[width=0.8\textwidth]{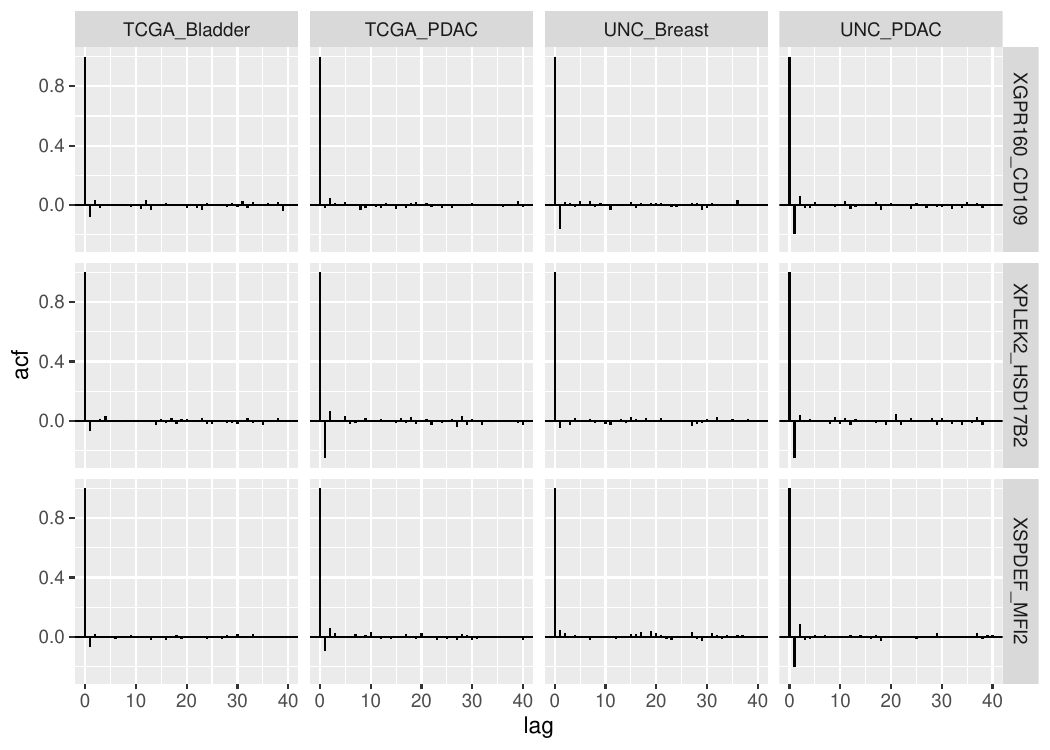}
    \caption{MCMC diagnostic plots for the basal best model results, created using the plot\_mcmc function. Top: Sample path plots for the random slopes of three TSP covariates. Bottom: Autocorrelation plots for the random slopes of three TSP covariates.}
    \label{fig:plotmcmc}
\end{figure}

The example code below specifies the names of three of TSP predictors with non-zero random effects across the studies and then uses the \code{plot\_mcmc} function to produce the sample path plots and autocorrelation plots for the corresponding posterior samples. Some plot aesthetics are adjusted using the \CRANpkg{ggplot2} package \citep{ggplot2book}. These sample path and autocorrelation plots can be seen in Figure \ref{fig:plotmcmc}.

\begin{example}
> TSP = c("XGPR160_CD109", "XSPDEF_MFI2", "XPLEK2_HSD17B2")
> plot_diag = plot_mcmc(object = fitB, plots = c("sample.path","autocorr"), 
+                       grps = "all", vars = TSP)
> library("ggplot2")
> plot_diag$sample_path + theme(axis.text.x = element_text(angle = 270))
> plot_diag$autocorr
\end{example}

\subsection{Optimization}
\label{sec:optimization}

Additional optimization control options can be passed to the \code{glmm} and
\code{glmmPen} functions using the \code{optim\_options} argument
and the \code{optimControl()} control structure. Some default settings
in \code{optimControl} depend on the family of the data or the number of
random effects. Descriptions of several of the main
\code{optimControl()} arguments and their defaults are listed
below. Disclaimer: Some optimization argument default values may be refined in future versions of the package if additional package testing suggests that changes could improve package performance (e.g.,~adjustments for certain data conditions or outcome families); please check the current \CRANpkg{glmmPen} documentation for the most up-to-date default information.

\code{sampler}: a character string specifying the sampling type used
in the E-step of the MCECM algorithm. The default sampler is ``stan'',
which requests the No-U-Turn Hamiltonian Monte Carlo sampling performed by
the \CRANpkg{rstan} package \citep{stan2020, stan2017}. We strongly
recommend using this sampling method due to its speed and efficiency.
Other options include ``random\_walk'', which requests the
Metropolis-within-Gibbs adaptive random walk sampler
\citep{adaptMCMC2009}, or ``independence'', which requests the
Metropolis-within-Gibbs independence sampler \citep{compstats2012}.

\code{var\_start}: either a character string ``recommend'' (default)
or a positive numeric value. This argument specifies the initial
starting variance of the random effects covariance matrix. If
\code{var\_start} is set to ``recommend'', the function fits a fixed
and random intercept only model using the \CRANpkg{lme4} package and sets
the starting variance to the random intercept variance multiplied by 2.
The random effects covariance matrix is initialized as a diagonal matrix
with the value of \code{var\_start} as the diagonal elements.

\code{var\_restrictions}: either a character string ``none'' (default) or the character string ``fixef''. This argument can be used to restrict which random effects are considered at the start of the algorithm. If this argument is set to ``none'', then all random effect predictors are initialized to have a non-zero variance in the random effect covariance matrix. If this argument is set to ``fixef'', then only the random effect predictors that are initialized to have non-zero fixed effects estimates during the fixed effects initialization procedure are given non-zero variances when initializing the random effect covariance matrix. In effect, this restricts predictors that are initialized with zero-valued fixed effects coefficients to not have random effects. See \textbf{glmmPen} simulation results utilizing this feature within the GitHub repository \url{https://github.com/hheiling/paper_glmmPen_RJournal}.
By using this restriction, the user assumes that predictors penalized out of the naive model do not have random effects. While this could be a strong assumption, using this restriction can be helpful in speeding up the algorithm by removing excessive random effects at the start of the variable selection procedure.

\code{conv\_EM}: a positive numeric value specifying the convergence
threshold for the EM algorithm. The argument \code{conv\_EM} specifies
the value of \(\epsilon\) in Equation~\ref{eqn:conv_EM}. The default
value is 0.0015.


\code{t}: a positive integer that specifies the value of \(t\) in the
EM algorithm convergence criteria specified in Equation~\ref{eqn:conv_EM}. The convergence critera is based on the average
Euclidean distance between the most recent coefficient estimate and the
coefficient estimate from \code{t} EM iterations back. Default value
is set to 2.

\code{mcc}: a positive integer indicating the number of times the EM
convergence criteria must be met before the algorithm is seen as having
converged (\code{mcc} short for `meet condition counter'). Default
value is set to 2, and \code{mcc} is restricted to be no less than 2.

\code{maxitEM}: The maximum number of EM iterations allowed by the
algorithm. When the default value of \code{NULL} is input,
\code{maxitEM} is set to a value that depends on the family type of
the data. For the Binomial and Poisson families, the default is set to
50. For the Gaussian family, the default is set to 100 (we have observed
that the Gaussian family data generally takes longer to converge).




Additional optimization parameters include M-step convergence parameters (\code{conv\_CD}, \code{maxit\_CD}), parameters specifying the number of posterior samples to acquire for the E-step throughout the algorithm (\code{nMC\_burnin}, \code{nMC\_start}, and \code{nMC\_max}), the number of posterior samples needed to calculate the log-likelihood (\code{M}), and the number of posterior samples to save for diagnostics (\code{nMC\_report}). Additional details about these parameters and their defaults can be found in the \CRANpkg{glmmPen} documentation of the function \code{optimControl}.

\section{Simulations}
\label{sec:simulations}

In this section, we present results from simulations in order to
examine the performance of our package.  We use the \CRANpkg{glmmPen} package to perform variable selection and examine the resulting fixed effects estimates as well as
the true and false positives for the fixed and random effects. All
simulations are performed using the default optimization settings
discussed in Section ``Optimization''.  While the performance of the original implementation of the pGLMM algorithm was demonstrated in \cite{rashid2020}, here we confirm the performance of the computational improvements made since then as well as newer features such as the pre-screening procedure.

\subsection{Simulation set-up}
\label{sec:setup}

We simulated binary responses from a logistic mixed-effects regression model with $p$ predictors. Of \(p\) total predictors, we assume that 2 predictors have truly non-zero fixed and random effects, and the other \(p-2\) predictors have zero-valued fixed and random effects. Our aim in the simulations was to select
the true predictors.

In these simulations, we consider the following situations: predictor dimensions of \(p=\{10,50\}\), sample size \(N=500\), number of groups \(K=\{5,10\}\), and standard deviation of the random effects \(\sigma=\{1,\sqrt{2}\}\). As discussed in Section ``Generalized linear mixed models'', we approximate the covariance matrix of the random effects as a diagonal matrix for these higher dimensions. We further consider the scenarios of moderate predictor effects, where the true fixed effects are \(\boldsymbol \beta = (0,1,1)^\top\).

For group \(k\), we generated the binary response \(y_{ki}\), \(i=1,...,n_k\) such that \(y_{ki} \sim Bernoulli(p_{ki})\) where \(p_{ki} = P(y_{ki} = 1 | \boldsymbol x_{ki}, \boldsymbol z_{ki}, \boldsymbol \alpha_k, \boldsymbol \theta) = \exp(\boldsymbol x_{ki}^\top \boldsymbol \beta + \boldsymbol z_{ki}^\top \boldsymbol \alpha_k) / \{1 + \exp(\boldsymbol x_{ki}^\top \boldsymbol \beta + \boldsymbol z_{ki}^\top \boldsymbol \alpha_k) \}\), and \(\boldsymbol \alpha_k \sim N_3(0, \sigma^2 \boldsymbol I_3)\).
The fixed effect coefficients were set to \(\boldsymbol \beta = (0,1,1)^\top\) (moderate predictor effects).
We also simulated imbalance in sample sizes between the groups. Of the \(N\) samples, \(N/3\) samples were given to study \(k=1\) and the remaining \(2N/3\) samples were evenly distributed among
the remaining studies. Each condition was evaluated using 100 total simulated datasets.

For individual \(i\) in group \(k\), the vector of predictors for the fixed effects was \(\boldsymbol x_{ki} = (1,x_{ki,1},...,x_{ki,p})^\top\), and we set the random effects \(\boldsymbol z_{ki} = \boldsymbol x_{ki}\), where \(x_{ki,j} \sim N(0,1)\) for \(j=1,...,p\).

Setting the input random effects equal to the fixed effects represents the worst-case scenario where we have no idea what predictors may or may not have random effects. This may be an extreme assumption; in many real-world scenarios, users will have reason to set the input random effects to a strict subset of the
fixed effects.

In all of these simulations, we use the default settings discussed earlier, which includes using the default $\boldsymbol \lambda_0$ and $\boldsymbol \lambda_1$ penalty sequences, BIC-ICQ for the selection criteria, pre-screening, and the MCP penalty. For all simulations, we performed the abbreviated two-stage grid search as described in Section ``Tuning parameter selection''. 
The results for these simulations are presented in Table \ref{tab:selectB1}. 
These results include the average coefficients, the average true positive and false positive percentages for both fixed and random effects, and the median time for the simulations to complete. The true positive percentages reflect the average percent of the true predictors included in the best models chosen by the BIC-ICQ model selection criteria, which should ideally be near 100\%. Likewise, the false positive percentages reflect the average percent of the false predictors included in the best models, which should ideally be near 0\%. All simulations were completed on the UNC Longleaf computing cluster (CPU Intel processors between 2.3Ghz and 2.5GHz).

\begin{table}[h!]

\centering
\begin{tabular}{ccccccccccc}
  \toprule
   $N$ & $p$ & $K$ & $\sigma$ & $\hat \beta_1$ & $\hat \beta_2$ & 
   \multicolumn{1}{p{1.2cm}}{\centering TP \% \\ Fixed}
   & \multicolumn{1}{p{1.2cm}}{\centering FP \% \\ Fixed}
   & \multicolumn{1}{p{1.35cm}}{\centering TP \% \\ Random}
   & \multicolumn{1}{p{1.35cm}}{\centering FP \% \\ Random}
   & \multicolumn{1}{p{1.3cm}}{\centering $T^{median}$ \\ (hours)}
   \\ 
  \midrule
  500 & 10 & 5  & 1        & 1.02 & 1.12 & 89.0 & 2.1 & 90.5 & 3.5 & 0.20 \\ 
      &    &    & $\sqrt{2}$ & 1.12 & 1.18 & 83.0 & 1.4 & 96.0 & 3.6 & 0.26 \\ 
      &    & 10 & 1        & 0.99 & 1.04 & 99.0 & 3.0 & 95.0 & 4.8 & 0.24 \\ 
      &    &    & $\sqrt{2}$ & 1.02 & 1.11 & 91.0 & 1.8 & 99.5 & 7.0 & 0.32 \\ 
  \midrule
  500 & 50 & 5  & 1        & 1.18 & 1.14 & 84.5 & 1.2 & 83.5 & 2.2 & 8.07 \\
      &    &    & $\sqrt{2}$ & 1.42 & 1.43 & 75.5 & 2.5 & 89.0 & 2.5 & 12.20 \\
      &    & 10 & 1        & 1.12 & 1.11 & 95.0 & 1.8 & 93.0 & 3.9 & 10.67 \\
      &    &    & $\sqrt{2}$ & 1.33 & 1.31 & 84.5 & 2.4 & 95.5 & 6.2 & 15.75 \\
   \bottomrule
\end{tabular}
\caption{Variable selection simulation results with moderate predictor effects (slopes equal to 1). Results include the estimated coefficients for true non-zero fixed effects, true positive (TP) percentages for fixed and random effects, false positive (FP) percentages for fixed and random effects, and the median time in hours for the algorithm to complete.}
\label{tab:selectB1}
\end{table}

By examining the simulation results, we can observe that the performance of the variable selection procedure in \code{glmmPen} is impacted by the underlying structure of the data. As the magnitude of the random effect variance increases, the true positive percentage of the fixed effects decreases and the true positive percentage of the random effects increases. Additionally, as the number of groups \(K\) increases, the true positive percentage of both the fixed and random effects increases. We see that as the dimension of the total number of predictors increases ($p=10$ to $p=50$), the true positive percentages of both the fixed and random effects decreases.
In regards to the run time, Table \ref{tab:selectB1} shows that increases in the number of groups and increases in the variance of the random effects generally increases the time for the algorithm to complete.

In simulations not presented in this paper, we saw that increases to the magnitude of the fixed effects (e.g.,~increasing the true slope to 2, see content in GitHub repository \url{https://github.com/hheiling/paper_glmmPen_RJournal} for details) increased the true positive fixed effects and generally decreased the true positive random effects.

\subsection{Pre-screening performance}
\label{sec:pre-screening}

The time it takes the package to complete the tuning parameter selection procedure
depends strongly on the number of random effects considered by the
algorithm. Therefore, the pre-screening procedure, which reduces the
number of random effects considered within the variable selection
algorithm, speeds up the algorithm. 
Table \ref{tab:prescB1} summarizes the performance of the pre-screening algorithm for the variable selection simulations described above. This table reports the average percent of true positive and false positive random effect predictors that remain under consideration within the variable selection procedure after the pre-screening step has completed.
The pre-screening settings were the default settings described in Section ``Software'', which include specifying \code{lambda.min.presc} = 0.01 for $p=10$ and \code{lambda.min.presc} = 0.05 for $p=50$ such that the minimum penalty on the random effects is \code{lambda.min.presc} $\times\lambda_{max}$. 
We note that there are currently no methods that are capable of scaling to the values of $q$ random effect predictors evaluated in our simulation for estimating and performing variable selection in GLMMs. 

\begin{table}[h!]

\centering
\begin{tabular}{cccccc}
  \toprule
   $N$ & $p$ & $K$ & $\sigma$  & TP \% & FP \% \\ 
  \midrule
  500 & 10 & 5  & 1         & 98.0 & 25.8 \\ 
      &    &    & $\sqrt{2}$ & 100.0 & 26.1 \\ 
      &    & 10 & 1         & 100.0 & 33.0 \\ 
      &    &    & $\sqrt{2}$  & 100.0 & 32.2 \\ 
  \midrule
  500 & 50 & 5  & 1         & 96.0 & 24.6 \\
      &    &    & $\sqrt{2}$  & 96.5 & 25.7 \\
      &    & 10 & 1         & 97.5 & 25.9 \\
      &    &    & $\sqrt{2}$  & 98.5 & 27.3 \\
   \bottomrule
\end{tabular}
\caption{Pre-screening results for variable selection simulations with moderate predictor effects (slopes equal to 1). Results include the true positive percentages and false positive percentages of the random effect predictors remaining after pre-screening.}
\label{tab:prescB1}
\end{table}



Using this higher penalty in the \(p=50\) simulations helps
reduce the false positive percentage of the random effects after pre-screening
and consequently helps speed up the time of the algorithm to complete.
However, we can see by comparing the \(p=50\) and \(p=10\) simulations
that this approach can also slightly decrease the true positive percentage. In general, increasing \code{lambda.min.presc} will help decrease the number of false positive non-zero random effects in the pre-screening step, but it may also decrease the number of true positive non-zero random effects. Decreasing \code{lambda.min.presc} will generally have the opposite effect. We also see that the true positive percentage for the selection of the random effects after pre-screening is generally higher when the magnitude of the true random effect variance is higher.

\section{Conclusion}
\label{sec:conclusion}

This paper introduces the R package \CRANpkg{glmmPen} for fitting
penalized generalized linear mixed models, including Binomial,
Gaussian, and Poisson models. The \CRANpkg{glmmPen} package's main advantage over other packages that estimate GLMMs is that it can perform
variable selection on the fixed and random effects simultaneously. The algorithm utilizes a Monte Carlo Expectation Conditional Minimization (MCECM) algorithm. Several
established MCMC sampling techniques are available for the E-step, and a
Majorization-Minimization coordinate descent algorithm is used in the
M-step. The package utilizes the established methods of \pkg{Stan} and
\CRANpkg{RcppArmadillo} to increase the computational efficiency of the
E-step and M-step, respectively. As a result, the \CRANpkg{glmmPen} package
can fit models with higher dimensions compared to other packages that
fit GLMMs, supporting models with 50 or more
fixed and random effects. 

The \CRANpkg{glmmPen} package employs several additional techniques to improve the speed of the algorithm. Such techniques include initialization of subsequent models with the
coefficients from the previous model fit and pre-screening to remove
unnecessary random effects.

The \CRANpkg{glmmPen} package has several attributes that make it user-friendly. For one, the package was designed to have an interface that is similar to the \CRANpkg{lme4} package, with which many users may be familiar. Additionally, the \CRANpkg{glmmPen}
package has several automated procedures that make it more convenient to use, including automated data-dependent initialization
of the random effect covariance matrix and
automated recommendations for the penalization parameters.

A unique aspect of the package is the calculation of the marginal
log-likelihood. The corrected arithmetic mean estimator (CAME)
calculation described by Pajor \citep{pajor2017} is relatively simple and fast to calculate, and we have found that it performs well when compared with the log-likelihood estimate used in the \CRANpkg{lme4} package (see content in the GitHub repository \url{https://github.com/hheiling/paper_glmmPen_RJournal}).
This marginal log-likelihood calculation allows the algorithm to perform tuning parameter selection using traditional BIC selection criterion as well as other BIC-derived selection criteria. This gives users the option to forgo calculating the BIC-ICQ selection criterion, which requires the minimal penalty model fit where a minimum penalty is applied to both the fixed and random effects.

In its current implementation at the time of this paper's publication, the \textbf{glmmPen} R package can apply to Binomial, Gaussian, and Poisson families with canonical links. In the future, we plan to extend the application of this package to survival data, to non-canonical links for the existing families, and to additional families such as the negative binomial family. 









\bibliography{heiling-glmmPen}

\address{Hillary M. Heiling\\
  University of North Carolina Chapel Hill\\
\email{hmheiling@gmail.com}
  }

\address{Naim U. Rashid\\
  University of North Carolina Chapel Hill\\
  \email{nur2@email.unc.edu}}

\address{Quefeng Li\\
  University of North Carolina Chapel Hill\\
  \email{quefeng@email.unc.edu}}

\address{Joseph G. Ibrahim\\
  University of North Carolina Chapel Hill\\
  \email{ibrahim@bios.unc.edu}}

\end{article}

\end{document}